\documentclass{tlp}

\usepackage{graphics}
\usepackage{times}
\usepackage{amssymb}
\usepackage[matrix,arrow]{xy}
\CompileMatrices

\newtheorem{theorem}{Theorem}
\newtheorem{proposition}[theorem]{Proposition}
\newtheorem{lemma}[theorem]{Lemma}
\newtheorem{corollary}[theorem]{Corollary}
\newtheorem{definition}{Definition}
\newtheorem{example}{Example}
\newtheorem{remark}{Remark}

\newcommand{\herbrand}{\mathcal{H}}
\newcommand{\assm}{{\it assm}}
\newcommand{\conc}{{\it conc}}
\newcommand{\mycenter}[1]{\hspace*{\fill}#1\hspace*{\fill}}

\newcommand{\na}[1]{{\sf #1}}

\author[Ralf Schweimeier and Michael Schroeder]
{Ralf Schweimeier and Michael Schroeder\\\\
Department of Computing, School of Informatics, City University\\
Northampton Square, London EC1V 0HB, UK\\\\
Department of Computer Science, Technische Universit\"at Dresden\\
01062 Dresden, Germany\\\\
\email{\{ralf,msch\}@soi.city.ac.uk}}
\title{A Parameterised Hierarchy of Argumentation Semantics for Extended Logic Programming and its Application to the Well-founded Semantics}

\shorttitle{A Hierarchy of Argumentation Semantics}

\jdate{Nov 2002}
\pubyear{2001}
\pagerange{\pageref{firstpage}--\pageref{lastpage}}
\doi{S1471068401001193}

\submitted{4 December 2002}
\revised{29 June 2003}
\accepted{12 September 2003}

\begin{document}

\maketitle

\begin{abstract}
  Argumentation has proved a useful tool in defining formal semantics
  for assumption-based reasoning by viewing a proof as a process in
  which proponents and opponents attack each others arguments by
  undercuts (attack to an argument's premise) and rebuts (attack to an
  argument's conclusion). In this paper, we formulate a variety of
  notions of attack for extended logic programs from combinations of
  undercuts and rebuts and define a general hierarchy of argumentation
  semantics parameterised by the notions of attack chosen by proponent
  and opponent. We prove the equivalence and subset relationships
  between the semantics and examine some essential properties
  concerning consistency and the coherence principle, which relates
  default negation and explicit negation.  Most significantly, we
  place existing semantics put forward in the literature in our
  hierarchy and identify a particular argumentation semantics for
  which we prove equivalence to the paraconsistent well-founded
  semantics with explicit negation, WFSX$_p$. 
  Finally, we present a general proof theory,
  based on dialogue trees, and show that it is sound and complete with
  respect to the argumentation semantics.
\end{abstract}


\newpage
\tableofcontents
\newpage

\section{Introduction}

Argumentation has attracted much interest in the area of Artificial
Intelligence. On the one hand, argumentation is an important way of
human interaction and reasoning, and is therefore of interest for
research into intelligent agents. Application areas include automated
negotiation via
argumentation~\cite{PSJ98:AgentsArguing,KSE98:Argument,Sch99:Argument}
and legal reasoning~\cite{PS97:Argument}.  On the other hand,
argumentation provides a formal model for various assumption based (or
non-monotonic, or default) reasoning
formalisms~\cite{BDKT97:Argument,CML00:LogicalModelsOfArgument}.  In
particular, various argumentation based semantics have been proposed
for logic programming with default
negation~\cite{BDKT97:Argument,Dun95:Argument}.

Argumentation semantics are elegant since they can be captured in an
abstract
framework~\cite{Dun95:Argument,BDKT97:Argument,%
Vre97:AbstractArgumentationSystems,JV99:RobustSemantics},
for which an elegant theory of attack, defence, acceptability, and
other notions can be developed, without recourse to the concrete
instance of the reasoning formalism at hand. This framework can then
be instantiated to various assumption based reasoning formalisms.
Similarly, a dialectical proof theory, based on dialogue trees,
can be defined for an abstract
argumentation framework, and then applied to any instance of such a
framework~\cite{SCG94:Dialectics,Dun95:Argument,JV99:DialecticSemantics}.

In general, an argument $A$ is a proof which may use a set of
defeasible assumptions.  Another argument $B$ may have a conclusion
which contradicts the assumptions or the conclusions of $A$, and
thereby $B$ {\em attacks} $A$.  There are two fundamental notions of such
attacks: undercut and rebut~\cite{Pol87:Defeasible,PS97:Argument} or equivalently {\em
ground-attack} and {\em reductio-ad-absurdum attack}
\cite{Dun93:ArgumentExplicit}.  We will use the terminology of
undercuts and rebuts. Both attacks differ in that an undercut attacks a
premise of an argument, while a rebut attacks a
conclusion. 

Given a logic program we can define an argumentation semantics by
iteratively collecting those arguments which are acceptable to a
proponent, i.e. they can be defended against all opponent attacks.  In
fact, such a notion of acceptability can be defined in a number of
ways depending on which attacks we allow the proponent and opponent to use.

Normal logic programs do not have negative conclusions, which means
that we cannot use rebuts. Thus both opponents can only launch
undercuts on each other's assumptions.
Various argumentation semantics have been defined for normal logic programs
\cite{BDKT97:Argument,Dun95:Argument,KT99:ComputingArgumentation}, 
some of which are equivalent to 
existing semantics such as the stable model semantics~\cite{GL88:StableModels}
or the well-founded semantics~\cite{GRS91:WFS}.

Extended logic programs
\cite{GL90:ClassicalNegation,AP96:WFSX,wag94a}, on the other hand,
introduce explicit negation, which states that a literal is explicitly
false. As a result, both undercuts and rebuts are possible forms of attack;
there are further variations depending on whether any kind of counter-attack
is admitted. A variety of argumentation semantics arise if one allows
one notion of attack as defence for the proponent, and another as attack
for the opponent.
Various argumentation semantics have been proposed for extended logic programs
\cite{Dun93:ArgumentExplicit,PS97:Argument,MA98:Argumentation,}. 
Dung has shown that a certain argumentation semantics is equivalent to
the answer set semantics~\cite{GL90:ClassicalNegation}, a generalisation
of the stable model semantics~\cite{GL88:StableModels}.
For the well-founded semantics with explicit negation,
WFSX~\cite{PA92:WFSX,AP96:WFSX}, there exists a \emph{scenario semantics} 
\cite{ADP93:ScenarioSemantics} which is similar to an argumentation semantics.
This semantics applies only to non-contradictory programs;
to our knowledge, no argumentation semantics has yet been found equivalent
to the \emph{paraconsistent} well-founded semantics with explicit negation,
WFSX$_p$~\cite{Dam96:Thesis,ADP95:LPsystem,AP96:WFSX}.

This paper makes the following contributions: we identify various
notions of attack for extended logic programs.  We set up a general
framework of argumentation semantics, parameterised on these notions
of attacks.  This framework is then used to classify notions of
justified arguments, and to compare them to the argumentation
semantics of 
\cite{Dun93:ArgumentExplicit} and 
\cite{PS97:Argument}, among others.  We examine some properties
of the different semantics, concerning consistency, and the coherence
principle which relates explicit and implicit negation.  One
particular argumentation semantics is then shown to be equivalent to
the paraconsistent well-founded semantics with explicit
negation~\cite{Dam96:Thesis}.  Finally, we
develop a general dialectical proof theory for the notions of
justified arguments we introduce, and show how proof procedures for
these proof theories can be derived. This paper builds upon an earlier
conference publication \cite{sch02a}, which reports initial findings,
while this article provides detailed coverage including all proofs and
detailed examples.

The paper is organised as follows: First we define arguments and
notions of attack and acceptability. Then we set up a framework for
classifying different least fixpoint argumentation semantics, based on
different notions of attack.  Section~\ref{sec:properties} examines
some properties (coherence and consistency) of these semantics.  In
Section~\ref{sec:wfsxarg}, we recall the definition of WFSX$_p$, and prove
the equivalence of an argumentation semantics and WFSX$_p$.  A general
dialectical proof theory for arguments is presented in
Section~\ref{sec:proof-theory}; we prove its soundness and
completeness and outline how a proof procedure for the proof theory
may be derived.

\section{Extended Logic Programming and Argumentation}
\label{sec:elp}

We introduce extended logic programming and summarise the definitions
of arguments associated with extended logic programs.
We identify various notions of attack between arguments,
and define a variety of semantics parametrised on these notions of attack.

Extended logic programming extends logic programming by two kinds of
negation: \emph{default negation} and \emph{explicit negation}.
The former allows the assumption of the falsity of a fact if there is
no evidence for this fact. Explicit negation, on the other hand,
allows to explicitly assert the falsity of a fact.

The default negation of a literal $p$, written $not~p$, states the
assumption of the falsity of $p$. The assumption $not~p$ is intended
to be true iff there is no evidence of $p$. Thus, the truth of $not~p$
relies on a lack of knowledge about $p$. 
An operational interpretation of default negation is given by
\emph{negation as failure} \cite{cla78}: 
the query $not~p$ succeeds iff the query $p$ fails.
Default negation is usually not allowed in the head of a rule: the
truth value of $not~p$ is defined in terms of $p$, and so there should
not be any other rules that define $not~p$.

Default negation thus gives a way of expressing a kind of negation,
based on a lack of knowledge about a fact. 
Sometimes, however, it is desirable to express the explicit knowledge
of the falsity of a fact. 
The explicit negation $\neg p$ of a literal $p$ states that $p$ is
known to be false. In contrast to default negation, an explicit
negation $\neg p$ is allowed in the head of a rule, and there is no
other way of deriving $\neg p$ except by finding an applicable rule
with $\neg p$ as its consequence.

Consider the following example
\footnote{Due to John McCarthy, first published in \cite{GL90:ClassicalNegation}}:
``A school bus may cross the railway tracks under the condition
that there is no approaching train.''
It may be expressed using default negation as
$$cross \gets not~train$$
This is a dangerous statement, however: assume that there is no knowledge about
an approaching train, e.g. because the driver's view is blocked. In
this case, the default negation $not~train$ is true, and we conclude
that the bus may cross.
Instead, it would be appropriate to demand the explicit knowledge that
there is no approaching train, as expressed using explicit negation:
$$cross \gets \neg train$$
The combination of default and explicit negation also allows for a
more cautious statement of positive facts: while the rule
$$\neg cross \gets train$$
states that the driver should not cross if there is a train approaching,
the rule
$$\neg cross \gets not~\neg train$$
states more cautiously that the driver should not cross if it has not
been established that there is no train approaching.
In contrast to the former rule, the latter rule prevents a driver from
crossing if there is no knowledge about approaching trains.

A connection between the two kind of negations may be made by
asserting the \emph{coherence principle}
\cite{PA92:WFSX,AP96:WFSX}: it states that whenever an
explicit negation $\neg p$ is true, then the default negation $not~p$
is also true. This corresponds to the statement that
if something is known to be false, then it should also be assumed to
be false.

\subsection{Arguments}
\label{subsec:arguments}

\begin{definition} 
  An {\em objective literal} is an atom $A$ or its explicit negation
  $\neg A$. We define $\neg \neg L = L$.
  A {\em default literal} is of the form $not~L$ where $L$
  is an objective literal. A {\em literal} is either an objective or a
  default literal.  
\\
  An {\em extended logic program} is a (possibly infinite) set of rules of
  the form 

    $L_0 \gets L_1,\dots,L_m, not~L_{m+1},\dots, not~L_{m+n} 
    (m, n \ge 0)$,
\\
  where each $L_i$ is an objective literal ($0\leq i \leq m+n$).
  For such a rule $r$, we call $L_0$ the {\em head} of the rule, 
  $head(r)$, and $L_1,\ldots,not~L_{m+n}$ the {\em body} of the rule, 
  $body(r)$.
  A rule with an empty body is called a {\em fact}, and we write
  $L_0$ instead of $L_0 \gets$.
\end{definition}

Our definition of an argument associated with an extended logic
program is based on~\cite{PS97:Argument}. 
Essentially, an argument is a partial proof, resting on a number of
{\em assumptions}, i.e.~a set of default literals.%
\footnote{In~\cite{BDKT97:Argument,Dun93:ArgumentExplicit}, an argument 
  {\em is} a set of assumptions; the two approaches are equivalent in 
  that there is an argument with a conclusion $L$ iff there is a set
  of assumptions from which $L$ can be inferred. See the discussion
  in~\cite{PS97:Argument}.}
Note that we do not consider priorities of rules, as used e.g.\
in~\cite{ant02,kak02,PS97:Argument,Bre96:DynamicPreferences,GSC98:Argument,%
Vre97:AbstractArgumentationSystems}.
Also, we do not distinguish between \emph{strict} rules, which may not
be attacked, and \emph{defeasible} rules, which may be attacked
\cite{PS97:Argument,SL92:MTDR,GSC98:Argument}.

\begin{definition} 
\label{def:argument}
Let $P$ be an extended logic program.  
An {\em argument} associated with $P$ is a finite sequence 
$A=[r_1,\dots r_n]$ of ground instances of
rules $r_i \in P$ such that
%
for every $1\le i \le n$, for every objective literal $L_j$ in
the body of $r_i$ there is a $k>i$ such that $head(r_k) = L_j$.
%
%
\\
A {\em subargument} of $A$ is a subsequence of $A$ which is an argument.
The head of a rule in $A$ is called a {\em conclusion} of $A$,
and a default literal $not~L$ in the body of a rule of $A$ is called an
{\em assumption} of $A$.
We write $\assm(A)$ for the set of assumptions and $\conc(A)$ for the 
set of conclusions of an argument $A$.

An argument $A$ with a conclusion $L$ is a {\em minimal argument for $L$} 
if there is no subargument of $A$ with conclusion $L$.
An argument is {\em minimal} if it is minimal for some literal $L$.
Given an extended logic program $P$, we denote the set of minimal arguments
associated with $P$ by $\mathit{Args}_P$.

\end{definition}

The restriction to minimal arguments (cf.~\cite{SL92:MTDR})
is not essential, but convenient,
since it rules out arguments constructed from several unrelated
arguments. Generally, one is interested in the conclusions of an
argument, and wants to avoid having rules in an argument which do not
contribute to the desired conclusion.
Furthermore, when designing a proof procedure to compute justified
arguments, one generally wants to compute only minimal arguments,
for reasons of efficiency.

\begin{example}
\label{ex:train}
Consider the following program:
$$\begin{array}{rcl}
  \neg cross & \gets & not~\neg train \\
  cross & \gets & \neg train \\
  train & \gets & see\_train \\
  \neg train & \gets & not~train, wear\_glasses \\
  wear\_glasses\\
\end{array}$$  
The program models the example from the introduction to this section.
A bus is allowed to cross the railway tracks if it is known that there
is no train approaching; otherwise, it is not allowed to cross.
A train is approaching if the driver can see the train,
and it is known that there is no train approaching if there is no
evidence of a train approaching, and the driver is wearing glasses.

There is exactly one minimal argument with conclusion $cross$:
$$[cross \gets \neg train; \neg train \gets not~train, wear\_glasses;
wear\_glasses]$$
It contains as subarguments the only minimal arguments for 
$\neg train$ and $wear\_glasses$:
$$[\neg train \gets not~train, wear\_glasses]$$
$$[wear\_glasses]$$
There is also exactly one minimal argument with conclusion $\neg cross$:
$$[\neg cross \gets not~\neg train]$$
There is no argument with conclusion $train$, because there is no rule
for $see\_train$.
\end{example}

\subsection{Notions of attack}
\label{subsec:attack}

There are two fundamental notions of attack: {\em undercut},
which invalidates an assumption of an argument, and
{\em rebut}, which contradicts a conclusion of an 
argument~\cite{Dun93:ArgumentExplicit,PS97:Argument}.
From these, we may define further
notions of attack, by allowing either of the two fundamental kinds of
attack, and considering whether any kind of counter-attack is allowed
or not. We will now formally define these notions of attack.

\begin{definition}
  \label{def:notionsofattack}
  Let $A_1$ and $A_2$ be arguments.
  \begin{enumerate}

  \item 
    $A_1$ {\em undercuts} $A_2$ if there is an objective literal
    $L$ such that $L$ is a conclusion of $A_1$ and $not~L$ is an 
    assumption of $A_2$.

  \item 
    $A_1$ {\em rebuts} $A_2$ if there is an objective literal $L$
    such that $L$ is a conclusion of $A_1$ and $\neg L$ is a
    conclusion of $A_2$.

  \item 
    $A_1$ {\em attacks} $A_2$ if $A_1$ undercuts or rebuts $A_2$.

  \item 
    $A_1$ {\em defeats} $A_2$ if 
    \begin{itemize}
    \item $A_1$ undercuts $A_2$, or 
    \item $A_1$ rebuts $A_2$ and $A_2$ does not undercut $A_1$.
    \end{itemize}    

  \item 
    $A_1$ {\em strongly attacks} $A_2$ if $A_1$ attacks $A_2$ and
    $A_2$ does not undercut $A_1$.

  \item 
    $A_1$ {\em strongly undercuts} $A_2$ if $A_1$ undercuts $A_2$
    and $A_2$ does not undercut~$A_1$.
  \end{enumerate}
\end{definition}

The notions of \emph{undercut} and \emph{rebut}, and hence \emph{attack} are
fundamental for extended logic
programs~\cite{Dun93:ArgumentExplicit,PS97:Argument}. The notion of
\emph{defeat} is used in~\cite{PS97:Argument}, along with a notion of
\emph{strict defeat}, i.e.\ a defeat that is not counter-defeated.
For arguments without priorities, rebuts are symmetrical, and
therefore strict defeat coincides with strict
undercut, i.e.\ an undercut that is not counter-undercut.
For this reason, we use the term {\em strong undercut} instead of 
{\em strict undercut}, and similarly define {\em strong attack} to be
an attack which is not counter-undercut.
We will use the following abbreviations for these notions of attack.
\na{r} for rebuts, 
\na{u} for undercuts, 
\na{a} for attacks, 
\na{d} for defeats, 
\na{sa} for strongly attacks, and
\na{su} for strongly undercuts.

\begin{example}
  Consider the program of example~\ref{ex:train}. There are the
  following minimal arguments:
  $$\begin{array}{ll}
    A: & [cross \gets \neg train; 
          \neg train \gets not~train, wear\_glasses;
          wear\_glasses] \\
    B: & [\neg cross \gets not~\neg train] \\
    C: & [\neg train \gets not~train, wear\_glasses] \\
    D: & [wear\_glasses] \\
  \end{array}$$
  The argument $A$ and $B$ rebut each other. 
  The subargument $C$ of $A$ also undercuts $B$, so $A$ also undercuts
  $B$. Therefore $A$ strongly attacks $B$, while $B$ does not strongly attack
  or defeat $A$.
\end{example}

\begin{example}
The arguments 
$[q\gets not~p]$
and
$[p\gets not~q]$
undercut each other. As a result, they do not strongly undercut each other.

The arguments $[p\gets not~q]$ and $[\neg p\gets not~r]$ do not
undercut each other, but strongly attack each other.

The argument $[\neg p\gets not~r]$ strongly undercuts $[p\gets not~
\neg p]$ and $[p\gets not~ \neg p]$ attacks - but does not defeat -
the argument $[\neg p\gets not~r]$.
\end{example}

These notions of attack define for any extended logic program a binary
relation on the set of arguments associated with that program.

\begin{definition}
  A {\em notion of attack} is a function $x$ which assigns to each
  extended logic program $P$ a binary relation $x_P$ on the set of
  arguments associated with $P$, 
  i.e.\ $x_P \subseteq \mathit{Args}_P\times \mathit{Args}_P$.  
  Notions of attack are partially ordered by defining
  $x \subseteq y \mbox{ ~iff~ } \forall P: x_P \subseteq y_P$
\end{definition}

\begin{description}
\item[Notation] We will use sans-serif font for the specific notions of
  attack introduced in Definition~\ref{def:notionsofattack} and
  their abbreviations: $\na{r}$, $\na{u}$, $\na{a}$, $\na{d}$,
  $\na{sa}$, and $\na{su}$. We will use $x, y, z, \ldots$ to denote 
  variables for notions of attacks. 
  Arguments are denoted by $A, B, C, \ldots$

  The term ``attack'' is somewhat overloaded: 1. it is the notion of
  attack \na{a} consisting of a rebut or an undercut; we use this
  terminology because it is standard in the
  literature~\cite{Dun93:ArgumentExplicit,PS97:Argument}.  2. in
  general, an attack is a binary relation on the set of arguments of a
  program; we use the term ``notion of attack''.  3. if the
  argumentation process is viewed as a dialogue between an proponent
  who puts forward an argument, and an opponent who tries to dismiss
  it, we may choose one notion of attack for the use of the proponent,
  and another notion of attack for the opponent.  In such a setting,
  we call the former notion of attack the ``defence'', and refer to
  the latter as ``attack'', in the hope that the meaning of the term
  ``attack'' will be clear from the context.
\end{description}

\begin{definition}
  Let $x$ be a notion of attack. Then 
  the {\em inverse} of $x$, denoted by $x^{-1}$, is defined as
  $x^{-1}_P = \{ (B,A) ~|~ (A,B) \in x_P \}$.
\end{definition}

In this relational notation,
Definition~\ref{def:notionsofattack} can be rewritten as 
$\na{a} = \na{u} \cup \na{r}$, $\na{d} = \na{u} \cup (\na{r} - \na{u}^{-1})$,
$\na{sa} = (\na{u} \cup \na{r}) - \na{u}^{-1}$, 
and $\na{su} = \na{u} - \na{u}^{-1}$.

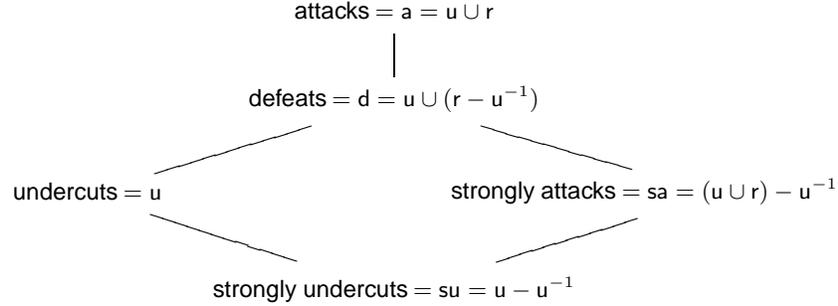
\begin{figure}
\hspace{-2em}
$$\xymatrixnocompile@C=-2em@R=4ex{
& \parbox{3.5cm}{\begin{center}\na{attacks} $= \na{a} = \na{u} \cup \na{r}$\end{center}\vspace{-1ex}} \ar@{-}[d] \\
& \parbox{5cm}{\vspace{-1.5ex}\begin{center}\na{defeats} $= \na{d} = \na{u} \cup (\na{r} - \na{u}^{-1})$\end{center}\vspace{-1ex}} 
\ar@{-}[dl] \ar@{-}[dr] \\
\parbox{3.5cm}{\vspace{-1.5ex}\begin{center}\na{undercuts} $= \na{u}$\end{center}\vspace{-1ex}} \ar@{-}[dr] && 
\hspace{-2cm}\parbox{6cm}{\vspace{-1.5ex}\begin{center}\na{strongly attacks} $= \na{sa} = (\na{u} \cup \na{r}) - \na{u}^{-1}$\end{center}\vspace{-1ex}} 
\ar@{-}[dl] \\
& \parbox{5.5cm}{\hspace{-2cm}\vspace{-1.5ex}\begin{center}\na{strongly undercuts} $= \na{su} = \na{u} - \na{u}^{-1}$\end{center}} \\
}$$
\caption{Notions of Attack}
\label{fig:notions-of-attack}
\end{figure}

\begin{proposition}
  The notions of attack of
  Definition~\ref{def:notionsofattack} are partially ordered according
  to the  diagram in Figure~\ref{fig:notions-of-attack}.
\end{proposition}

\begin{proof}
  A simple exercise, using the set-theoretic laws $A-B \subseteq A
  \subseteq A \cup C$ and $(A \cup B) - C = (A - C) \cup (B - C)$ 
  (for any arbitrary sets $A$, $B$, and $C$).
\end{proof}

As mentioned above, we will work with notions of attack as examined in
previous literature. Therefore Figure~\ref{fig:notions-of-attack}
contains the notions of
\na{undercut} \cite{Dun93:ArgumentExplicit,PS97:Argument}, 
\na{attack}  \cite{Dun93:ArgumentExplicit,PS97:Argument}, 
\na{defeat} \cite{PS97:Argument}, 
\na{strong undercut}  \cite{PS97:Argument}, 
and \na{strong attack} as  an intermediate 
notion between \na{strongly undercuts} and \na{defeats}.
All of these notions of attack are extensions of \na{undercuts}.
The reason is
that undercuts are asymmetric, i.e. for two arguments $A$, $B$, $A
\na{u} B$ does not necessarily imply $B \na{u} A$. 
Rebuts, on the other hand, are symmetric, i.e.  $A \na{r} B$ implies
$B \na{r} A$.  As a consequence, rebuts on their own always lead to a
``draw'' between arguments. There is, however, a lot of work on
priorities between arguments
\cite{ant02,kak02,PS97:Argument,Bre96:DynamicPreferences,%
GSC98:Argument,Vre97:AbstractArgumentationSystems}, which implies that
rebuts become asymmetric and therefore lead to more interesting
semantics. But the original, more basic approach does not consider
this extension, and hence undercuts play the prime role and notions of
attack mainly based on rebuts, such as $\na{r}$ or $\na{r-u^{-1}}$,
are not considered.

The following example shows that the inclusions in
Figure~\ref{fig:notions-of-attack} are strict.

\begin{example}
Consider the following program:

\begin{center}
  $\begin{array}{rcl}
    p & \gets & not~\neg p \\
    p & \gets & not~q \\
    \neg p & \gets & not~r \\
    q & \gets & not~p \\
    \neg~q & \gets & not~s \\
  \end{array}$
\end{center}

  \noindent
  It has the minimal arguments
  $\{ [p \gets not~\neg p], [p \gets not~q], [\neg p \gets not~r],
  [q \gets not~p], [\neg~q \gets not~s] \}$.
  The arguments $[p \gets not~q]$ and $[q \gets not~p]$ undercut
  (and hence defeat) each other, but they do not strongly undercut
  or strongly attack each other. 
  The arguments $[q \gets not~r]$ and $[\neg q \gets not~s]$
  strongly attack (and hence defeat) each other, but they do not
  undercut each other.
  The argument $[p \gets not~\neg p]$ attacks $[\neg p \gets not~r]$,
  but it does not defeat it, because $[\neg p \gets not~r]$
  (strongly) undercuts $[p \gets not~\neg p]$.
\end{example}

\subsection{Acceptability and justified arguments}
\label{subsec:acceptable}

Given the above notions of attack, we define acceptability of an
argument. Basically, an argument is acceptable if it can be defended
against any attack.  Our definition of acceptability is parametrised
on the notions of attack allowed for the proponent and the opponent.

Acceptability forms  the basis for our argumentation semantics,
which is defined as the least fixpoint of a function, which collects
all acceptable arguments
\cite{Pol87:Defeasible,SL92:MTDR,PS97:Argument,Dun93:ArgumentExplicit}.
The {\em least} fixpoint is of particular interest, because it
provides a canonical fixpoint semantics and it can be constructed
inductively.

Because the semantics is based on parametrised acceptability, we
obtain a uniform framework for defining a variety of argumentation
semantics for extended logic programs. It can be instantiated to a
particular semantics by choosing one notion of attack for the
opponent, and another notion of attack as a defence for the proponent.
The uniformity of the definition makes it a convenient framework for
comparing different argumentation semantics.


\begin{definition}
  Let $x$ and $y$ be notions of attack. Let $A$ be an argument, and
  $S$ a set of arguments. Then $A$ is {\em $x/y$-acceptable wrt.\ $S$}
  if for every argument $B$ such that $(B,A) \in x$ there exists an
  argument $C \in S$ such that $(C,B) \in y$.
\end{definition}

Based on the notion of acceptability, we can then define a fixpoint
semantics for arguments.

\begin{definition}
\label{def:just}
  Let $x$ and $y$ be notions of attack, and $P$ an extended logic
  program. The operator 
  $F_{P, x/y}:\mathcal{P}(\mathit{Args}_P) \rightarrow \mathcal{P}(\mathit{Args}_P)$
  is defined as
  $$F_{P, x/y}(S) = \{ A ~|~ A \mbox{ is $x/y$-acceptable wrt.\ $S$} \}$$
  We denote the least fixpoint of $F_{P, x/y}$ by $J_{P, x/y}$.
  If the program $P$ is clear from the context, we omit the subscript $P$.
  An argument $A$ is called {\em $x/y$-justified} if $A \in J_{x/y}$;
  an argument is called {\em $x/y$-overruled} if it is attacked by an
  $x/y$-justified argument; and an argument is called {\em $x/y$-defensible}
  if it is neither $x/y$-justified nor $x/y$-overruled.
\end{definition}

Note that this definition implies that the logic associated with
justified arguments is 3-valued, with justified arguments corresponding
to {\em true} literals, overruled arguments to {\em false} literals,
and defensible arguments to {\em undefined} literals.
We could also consider arguments which are both justified and overruled;
these correspond to literals with the truth value {\em overdetermined}
of Belnap's four-valued logic~\cite{Bel77:FourValued}.

\begin{proposition}
For any program $P$, the operator $F_{P,x/y}$ is monotone.
By the Knaster-Tarski fixpoint
theorem~\cite{Tar55:Fixpoint,Bir67:LatticeTheory}, $F_{P,x/y}$ has a
least fixpoint. It can be constructed by transfinite induction as follows:
\begin{quote}
  $\begin{array}{llll}
    J^0_{x/y} & \!\!\!:=\!\!\! & \emptyset \\
    J^{\alpha+1}_{x/y} & \!\!\!:=\!\!\! & F_{P, x/y}(J^\alpha_{x/y}) &
    \mbox{for $\alpha \! + \! 1$ a successor ordinal} \\
    J^\lambda_{x/y} & \!\!\!:=\!\!\! & \bigcup_{\alpha <
      \lambda}J^\alpha_{x/y} &
    \mbox{for $\lambda$ a limit ordinal} \\
\end{array}$
\end{quote}
Then there exists a least ordinal $\lambda_0$ such that
$F_{x/y}(J^{\lambda_0}_{x/y}) = J^{\lambda_0}_{x/y} =: J_{x/y}$.
\end{proposition}

\begin{proof}
  Let $S_1 \subseteq S_2$, and $A \in F_{P,x/y}$, 
  i.e.~$A$ is $x/y$-acceptable wrt.~$S_1$, i.e.~every $x$-attack against
  $A$ is $y$-attacked by an argument in $S_1$. Then $A$ is also
  $x/y$-acceptable wrt.~$S_2$, because $S_1 \subseteq S_2$, i.e.~$S_2$
  contains more arguments to defend $A$.
\end{proof}

Note that our general framework encompasses some well-known argumentation
semantics for extended logic programs: 
Dung's grounded semantics~\cite{Dun93:ArgumentExplicit} is $J_{\na{a}/\na{u}}$.
Prakken and Sartor's argumentation semantics~\cite{PS97:Argument}, 
without priorities or strict rules is $J_{\na{d}/\na{su}}$.
If we regard explicitly negated literals $\neg L$ as new atoms, unrelated
to the positive literal $L$, then we can apply the well-founded
argumentation semantics of~\cite{BDKT97:Argument,KT99:ComputingArgumentation} 
to extended logic programs, and obtain $J_{\na{u}/\na{u}}$.

  \begin{table}[t]
    \begin{center}
      $\begin{array}{l||c|c|c|c|c|c|c}
        & \na{a}/x & \na{d}/x & 
        \parbox{3em}{$\na{u}/\na{u} =$ $\na{u}/\na{su}$} & 
        \parbox{4em}{$\na{u}/\na{a} =$ $\na{u}/\na{d} =$ $\na{u}/\na{sa}$} & 
        \parbox{4em}{$\na{sa}/\na{sa} =$ $\na{sa}/\na{su}$} &
        \parbox{4em}{$\na{sa}/\na{a} =$ $\na{sa}/\na{d} =$ $\na{sa}/\na{u}$} & \na{su}/x \\ \hline\hline
        1 & \emptyset & [s] & [s] & [s] &
        \parbox[c]{5em}{$[p~\gets~not~q]$, $[s]$} & 
        \parbox[c]{5em}{$[p~\gets~not~q]$, $[s]$} & 
        \parbox[c]{5em}{$[p~\gets~not~q]$, $[q~\gets~not~p]$, $[s]$} 
        \\ \hline
        2 & \emptyset & \emptyset & [\neg q \gets not~r] & 
        [\neg q \gets not~r] & \emptyset & [\neg q \gets not~r] & 
        [\neg q \gets not~r] \\ \hline
        3 & \emptyset & \emptyset & \emptyset & [p \gets not~q] &
        \emptyset & \emptyset & \emptyset \\ \hline
        4 & \emptyset & \emptyset & \emptyset & \emptyset & \emptyset & 
        \emptyset & \emptyset
      \end{array}$
      \caption{Computing justified arguments --
        the $n$-th row shows the justified arguments 
        added at the $n$-th iteration}
      \label{tab:justified}
    \end{center}
  \end{table}

\begin{example}
Consider the following program $P$:

\begin{center}
  $\begin{array}{rcl}
    p & \gets & not~q \\
    q & \gets & not~p \\
    \neg q & \gets & not~r \\
    r & \gets & not~s \\
    s \\
    \neg s & \gets & not~s \\
  \end{array}$
\end{center}

  Table~\ref{tab:justified} shows the computation of justified
  arguments associated with $P$.
  The columns show  various combinations $x/y$ of attack/defence,
  and a row $n$ shows those arguments $A$ that get added at iteration
  stage $n$, i.e. $A \in J^n_{P, x/y}$ and 
  $A \not\in J^{n-1}_{P, x/y}$.

  The set of arguments associated with $P$ is
  $\{ [p \gets not~q], [q \gets not~p], [\neg q \gets not~r],
      [r \gets not~s], [s], [\neg s \gets not~s] \}$.

  All arguments are undercut by another argument, except $[s]$;
  the only attack against $[s]$ is a rebut by $[\neg s \gets not~s]$,
  which is not a defeat. Thus, $[s]$ is identified as a justified argument
  at stage $0$ in all semantics, except if \na{attacks} 
  is allowed as an attack.
  In the latter case, no argument is justified at stage $0$, hence the
  set of justified arguments $J_{\na{a}/x}$ is empty.

\end{example}

\section{Relationships between Notions of Justifiability}
\label{sec:relationship}

The definition of justified arguments provides a variety of 
semantics for extended logic programs, depending on which
notion of attack $x$ is admitted to attack an argument,
and which notion of attack $y$ may be used as a defence.

This section is devoted to an analysis of the relationship 
between the different notions of justifiability, leading to
a hierarchy of notions of justifiability illustrated in 
Figure~\ref{fig:hierarchy}.

\subsection{Equivalence of argumentation semantics}
\label{sec:eq}

We will prove a series of theorems, which show that some of
the argumentation semantics defined above are subsumed by others,
and that some of them are actually equivalent.
Thus, we establish a hierarchy of argumentation semantics, which 
is illustrated in Figure~\ref{fig:hierarchy}.

First of all, it is easy to see that the least fixpoint increases if
we weaken the attacks or strengthen the defence.
 
\begin{theorem}
  \label{thm:subset-just}
  Let $x' \subseteq x$ and $y \subseteq y'$ be notions of attack, then
  $J_{x/y} \subseteq J_{x'/y'}$.
\end{theorem}

\begin{proof}
  See \ref{appendix}.
\end{proof}

Theorem~\ref{thm:strong-nonstrong} states that it does not make a
difference if we allow only the strong version of the defence. This is
because an argument need not defend itself on its own, but it may rely
on other arguments to defend it.

\begin{theorem}
  \label{thm:strong-nonstrong}
  Let $x$ and and $y$ be notions of attack such that
  $x \supseteq \na{undercuts}$, and let
  $sy = y - \na{undercuts}^{-1}$.  
  Then $J_{x/y} = J_{x/sy}$.
\end{theorem}

\begin{proof} 
  Informally, every $x$-attack $B$ to an $x/y$-justified argument $A$ is
  $y$-defended by some $x/sy$-justified argument $C$ (by induction). 
  Now if $C$ is {\em not} a $sy$-attack, then it is undercut by $B$,
  and because $x \supseteq \na{undercuts}$ and 
  $C$ is justified, there exists a {\em strong} defence
  for $C$ against $B$, which is also a defence of the original
  argument $A$ against $C$.\\[2ex]
  The formal proof is by transfinite induction.
  By Theorem~\ref{thm:subset-just}, we have $J_{x/sy} \subseteq J_{x/y}$.
  We prove the inverse inclusion by showing that
  for all ordinals $\alpha$:
  $J_{x/y}^\alpha \subseteq J_{x/sy}^\alpha$,
  by transfinite induction on $\alpha$.
  See \ref{appendix} for the detailed proof.
\end{proof}

In particular, the previous Theorem states that undercut and strong
undercut are equivalent as a defence, as are attack and strong attack.
This may be useful in an implementation, where we may use the stronger
notion of defence without changing the semantics, thereby decreasing
the number of arguments to be checked.
The following Corollary shows that because defeat lies between attack
and strong attack, it is equivalent to both as a defence.

\begin{corollary}
  \label{cor:xa-xsa-just}
  Let $x$ be a notion of attack such that $x \supseteq \na{undercuts}$. Then
  $J_{x/\na{a}} = J_{x/\na{d}} = J_{x/\na{sa}}$.
\end{corollary}

\begin{proof}
  It follows from 
  Theorems~\ref{thm:subset-just} and~\ref{thm:strong-nonstrong} that
  $J_{x/\na{sa}} \subseteq J_{x/\na{d}} \subseteq J_{x/\na{a}} = J_{x/\na{sa}}$.
\end{proof}

The following theorem states that defence with \na{undercuts} is 
equally strong as one with \na{defeats} or with \na{attacks},
provided the opponent's permitted attacks include at least the
\na{strong attacks}.

\begin{theorem}
  \label{thm:xu-xa-just}
  Let $x$ be a notion of attack such that 
  $x \supseteq$ \na{strongly attacks}. Then
  $J_{x/\na{u}} = J_{x/\na{d}} = J_{x/\na{a}}$.
\end{theorem}

\begin{proof}
  It is sufficient to show that $J_{x/\na{a}} \subseteq J_{x/\na{u}}$.
  Then by Theorem~\ref{thm:subset-just}, 
  $J_{x/\na{u}} \subseteq J_{x/\na{d}} \subseteq J_{x/\na{a}} = J_{x/\na{u}}$.\\[1ex]
  Informally, every $x$-attack $B$ to a $x/\na{a}$-justified argument $A$
  is attacked by some
  $x/\na{u}$-justified argument $C$ (by induction). 
  If $C$ is a rebut, but not an
  undercut, then because $B$ strongly attacks $C$, and because
  $x \supseteq$ \na{strongly attacks},
  there must have been an argument defending $C$ by
  undercutting $B$, thereby also defending $A$ against $B$.\\[1ex]
  We prove by transfinite induction that for all ordinals $\alpha$:
  $J^\alpha_{x/\na{a}} \subseteq J^\alpha_{x/\na{u}}$.
  See \ref{appendix} for the detailed proof.
\end{proof}

In analogy to Theorem~\ref{thm:xu-xa-just}, strong undercuts are an 
equivalent defence to strong attacks if the allowed attacks are
strong attacks.

\begin{theorem}
  \label{thm:sasu-sasa-just}
  $J_{\na{sa}/\na{su}} = J_{\na{sa}/\na{sa}}$
\end{theorem}

\begin{proof}
  The proof is similar to the proof of Theorem~\ref{thm:xu-xa-just}.
  See \ref{appendix}.
\end{proof}

\begin{theorem}
  \label{thm:sua-sud-just}
  $J_{\na{su}/\na{a}} = J_{\na{su}/\na{d}}$
\end{theorem}

\begin{proof}
  By Theorem~\ref{thm:subset-just}, 
  $J_{\na{su}/\na{d}} \subseteq J_{\na{su}/\na{a}}$.\\[1ex]
  We now show the inverse inclusion.  Informally, every strong
  undercut $B$ to a $\na{su}/\na{a}$-justified argument $A$ is attacked by some
  $\na{su}/\na{d}$-justified argument $C$ (by induction).  If $C$ does not
  defeat $A$, then there is some argument $D$ defending $C$ by
  defeating $B$, thereby also defending $A$ against $B$.\\[1ex]
  Formally, we show that
  for all ordinals $\alpha$:
  $J_{\na{su}/\na{a}}^\alpha \subseteq J_{\na{su}/\na{d}}^\alpha$,
  by transfinite induction on $\alpha$.
  See \ref{appendix} for the detailed proof.
\end{proof}

These results are summarised in a hierarchy of argumentation semantics
in Theorem~\ref{thm:hierarchy} and Figure~\ref{fig:hierarchy}.

\subsection{Distinguishing argumentation semantics}
\label{sec:ex}

The previous section showed equality and subset relationships for a
host of notions of justified arguments. In this section we complement
these positive findings by negative findings stating for which
semantics there are no subset relationships. We prove these negative
statements by giving counter-examples distinguishing various notions of
justifiability.

The first example shows that, in general, allowing only strong forms
of attack for the opponent leads to a more credulous semantics,
because in cases where only non-strong attacks exist, every argument
is justified.

\begin{example}\hspace{0cm}
  \label{ex:loop}
Consider the following program:
\begin{center}
    $\begin{array}{rcl}
      p & \gets & not~q \\
      q & \gets & not~p \\
    \end{array}$
\end{center}
  For any notion of attack $x$, we have
  $J_{\na{su}/x} = J_{\na{sa}/x} = \{ [p \gets not~q], [q \gets not~p] \},$
  because there is no strong undercut or strong attack to any
  of the arguments.
  However, 
  $J_{\na{a}/x} = J_{\na{d}/x} = J_{\na{u}/x} = \emptyset,$
  because every argument is undercut (and therefore defeated and
  attacked).

  Thus, in general, $J_{s/x} \not\subseteq J_{w/y}$, for
  $s \in \{\na{su}, \na{sa}\}$, $w \in \{\na{a}, \na{u}, \na{d}\}$, and any notions of 
  attack $x$ and $y$.
\end{example}

The following example shows that some interesting properties need not
hold for all argumentation semantics: a fact (i.e. a rule with an
empty body) need not necessarily lead to a justified argument;
this property distinguishes Dung's~\cite{Dun93:ArgumentExplicit} and
Prakken and Sartor's~\cite{PS97:Argument} semantics from most of the
others.

\begin{example}\hspace{0em}
  \label{ex:coherence}
Consider the following program:
\begin{center}
    $\begin{array}{rcl}
    p & \gets & not~q \\
    q & \gets & not~p \\
    \neg p \\
  \end{array}$
\end{center}
  Let $x$ be a notion of attack. Then 
  $J_{\na{d}/x} = J_{\na{a}/x} = \emptyset,$
  because every argument is defeated (hence attacked).
  $J_{\na{sa}/\na{su}} = J_{\na{sa}/\na{sa}} = \{ [q \gets not~p] \},$
  because $[q \gets not~p]$ is the only argument which is not
  strongly attacked, but it does not strongly attack any other
  argument.
  $J_{\na{u}/\na{su}} = J_{\na{u}/\na{u}} = \{ [\neg p] \},$
  because there is no undercut to $[\neg p]$, but 
  $[\neg p]$ does not undercut any other argument.
  $J_{\na{u}/\na{a}} = \{ [\neg p], [q \gets not~p] \},$
  because there is no undercut to $[\neg p]$, and the undercut
  $[p \gets not~p]$ to $[q \gets not~p]$ is attacked by 
  $[\neg p]$. We also have
  $J_{\na{sa}/\na{u}} = \{ [\neg p], [q \gets not~p] \},$
  because $[q \gets not~p]$ is not strongly attacked, and the strong
  attack $[p \gets not~q]$ on $[\neg p]$ is undercut by
  $[q \gets not~p]$.

Thus, in general, $J_{\na{u}/x} \not\subseteq J_{\na{d}/x}$,
$J_{\na{u}/x} \not\subseteq J_{\na{a}/x}$,
$J_{\na{sa}/sx} \not\subseteq J_{\na{u}/y}$ (where $sx \in \{ \na{su}, \na{sa} \}$ and
$y \in \{ \na{u}, \na{su} \}$), and
$J_{\na{u}/y} \not\subseteq J_{\na{sa}/sx}$ (where $sx \in \{ \na{su}, \na{sa} \}$ and
$y \in \{ \na{u}, \na{a}, \na{d}, \na{su}, \na{sa}\}$).
\end{example}

The following example is similar to the previous example,
except that all the undercuts are strong, whereas in the previous
example there were only non-strong undercuts.

\begin{example}\hspace{0cm}
  \label{ex:coherence-big}
Consider the following program:
\begin{center}
    $\begin{array}{rcl}
      p & \gets & not~q \\
      q & \gets & not~r \\
      r & \gets & not~s \\
      s & \gets & not~p \\
      \neg p \\
    \end{array}$
\end{center}
  Let $x$ be a notion of attack.
  Then $J_{\na{sa}/x} = \emptyset,$
  because every argument is strongly attacked.

  $J_{\na{su}/\na{u}} = J_{\na{su}/\na{su}} = \{ [\neg p] \},$ because
  all arguments except $[\neg p]$ are strongly undercut, but $[\neg
  p]$ does not undercut any argument.  And $J_{\na{u}/\na{a}} =
  J_{\na{su}/\na{sa}} = J_{\na{su}/\na{a}} = \{ [\neg p], [q \gets
  not~r], [s \gets not~p] \}$, because $[\neg p]$ is not undercut, and it
  defends $[s \gets not~p]$ against the strong undercut $[p \gets not~q]$ (by
  rebut), and in turn, $[s \gets not~p]$ defends $[q \gets not~r]$ against the
  strong undercut $[r \gets not~s]$ (by strong undercut).

  Thus, $J_{\na{u}/\na{a}} \not\subseteq J_{\na{su}/y}$,
  $J_{\na{su}/\na{sa}} \not\subseteq J_{\na{su}/y}$, and
  $J_{\na{su}/\na{a}} \not\subseteq J_{\na{su}/y}$, for $y \in \{
  \na{u}, \na{su} \}$.
\end{example}

The following example shows that in certain circumstances,
non-strong defence allows for more justified arguments than
strong defence.

\begin{example}\hspace{0cm}
  \label{ex:loop-plus}
Consider the following program:
\begin{center}
    $\begin{array}{rcl}
      p & \gets & not~q \\
      q & \gets & not~p \\
      r & \gets & not~p \\
    \end{array}$
\end{center}
  Let $x$ be a notion of attack. Then
  $J_{\na{u}/x} = J_{\na{d}/x} = J_{\na{a}/x} = \emptyset,$
  because every argument is undercut.
  $J_{\na{su}/\na{su}} = J_{\na{su}/\na{sa}} = J_{\na{sa}/\na{su}} = J_{\na{sa}/\na{sa}} = 
    \{ [p \gets not~q], [q \gets not~p] \}$ :
  In these cases, the strong attacks are precisely the strong undercuts;
  the argument $[r \gets not~p]$ is not justified, because the 
  strong undercut $[p \gets not~q]$ is undercut, but not strongly 
  undercut, by $[q \gets not~p]$. And finally, 
  $J_{\na{su}/\na{u}} = J_{\na{su}/\na{a}} = J_{\na{sa}/\na{u}} = J_{\na{sa}/\na{a}} = 
    \{ [p \gets not~q], [q \gets not~p], [r \gets not~p] \}$ :
  Again, undercuts and attacks, and strong undercuts and strong
  attacks, coincide; but now $[r \gets not~p]$ is justified,
  because non-strong undercuts are allowed as defence.

  Thus, in general, $J_{x/\na{u}} \not\subseteq J_{x/\na{su}}$ and
  $J_{x/\na{a}} \not\subseteq J_{x/\na{sa}}$, where $x \in \{ \na{su}, \na{sa} \}$.
\end{example}

The following example distinguishes the argumentation semantics
of Dung~\cite{Dun93:ArgumentExplicit} and 
Prakken and Sartor~\cite{PS97:Argument}.

\begin{example}\hspace{0cm}
  \label{ex:defeat}
Consider the following program:
\begin{center}
    $\begin{array}{rcl}
      p & \gets & not~\neg p \\
      \neg p \\
    \end{array}$
\end{center}
  Then $J_{\na{a}/x} = \emptyset$, because both arguments attack each
  other, while $J_{\na{d}/x} = \{[\neg p]\}$, because $[\neg p]$ defeats
  $[p \gets not~\neg p]$, but not vice versa.

Thus, $J_{\na{d}/x} \not\subseteq J_{\na{a}/x}$.
\end{example}

The final example shows that if we do not allow any rebuts as attacks,
then we obtain a strictly more credulous semantics.

\begin{example}\hspace{0cm}
  \label{ex:loop-neg}
Consider the following program:
\begin{center}
    $\begin{array}{rcl}
      \neg p & \gets & not~q \\
      \neg q & \gets & not~p \\
      p \\
      q \\
  \end{array}$
\end{center}
  Let $x$ be a notion of attack. Then
  $J_{\na{sa}/x} = J_{\na{d}/x} = J_{\na{a}/x} = \emptyset,$
  because every argument is strongly attacked (hence defeated and attacked),
  while
  $J_{\na{u}/x} = J_{\na{su}/x} = \{ [p], [q] \}.$

  Thus, in general, $J_{v/x} \not\subseteq J_{w/y}$, where $v \in \{
  \na{u}, \na{su} \}$, $w \in \{ \na{a}, \na{d}, \na{sa} \}$, and $x$
  and $y$ are any notions of attack.
\end{example}

\subsection{A hierarchy of argumentation semantics}

We now summarise the results of this section, establishing a complete
hierarchy of argumentation semantics, parametrised on a pair of 
notions of attack $x/y$ where $x$ stands for the attacks on an argument,
and $y$ for the possible defence.
We locate in this hierarchy
the argumentation semantics of Dung~\cite{Dun93:ArgumentExplicit}
and Prakken and Sartor~\cite{PS97:Argument}, as well as the 
well-founded semantics for normal logic programs~\cite{GRS91:WFS}.
In Section~\ref{sec:wfsxarg} we will show that the paraconsistent
well-founded semantics with explicit negation,
WFSX$_p$~\cite{Dam96:Thesis}, 
can also be found in our hierarchy.
As a corollary, we obtain precise relationships between these well-known
semantics and our argumentation semantics.

\begin{figure}[htbp]
  \begin{center}
  \footnotesize $\xymatrix@C=-2em@R=3ex{ &&
  \na{su}/\na{a} = \na{su}/\na{d} \\ & \na{su}/\na{u} \ar@{-}[ru] &&
  \na{su}/\na{sa} \ar@{-}[lu] \\ \na{sa}/\na{u} = \na{sa}/\na{d} =
  \na{sa}/\na{a} \ar@{-}[ru] && \na{su}/\na{su} \ar@{-}[lu]
  \ar@{-}[ru] && \na{u}/\na{a} = \na{u}/\na{d} = \na{u}/\na{sa}
  \ar@{-}[lu] \\ & \na{sa}/\na{su} = \na{sa}/\na{sa} \ar@{-}[lu]
  \ar@{-}[ru] && \na{u}/\na{su} = \na{u}/\na{u} \ar@{-}[lu]
  \ar@{-}[ru] \\ && \na{d}/\na{su} = \na{d}/\na{u} = \na{d}/\na{a} =
  \na{d}/\na{d} = \na{d}/\na{sa} \ar@{-}[lu] \ar@{-}[ru] \\ &&
  \na{a}/\na{su} = \na{a}/\na{u} = \na{a}/\na{a} = \na{a}/\na{d} =
  \na{a}/\na{sa} \ar@{-}[u] \\ }$ 

  \caption{Hierarchy of Notions of Justifiability} 
  \label{fig:hierarchy} 
  \end{center}
\end{figure}
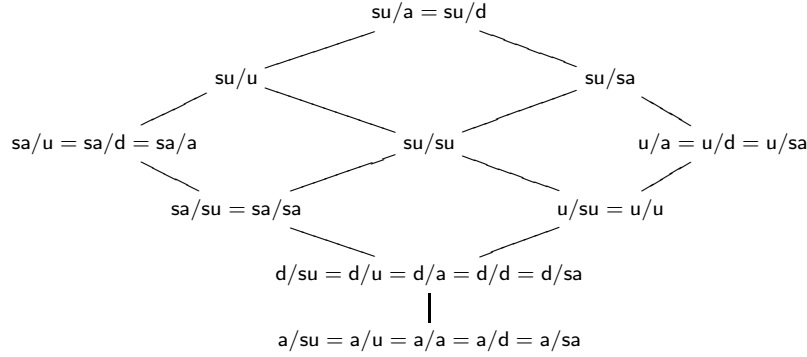

\begin{theorem}
  \label{thm:hierarchy}
  The notions of justifiability are ordered (by set inclusion)
  according to the diagram in Figure~\ref{fig:hierarchy},
  where $x/y$ lies below $x'/y'$ iff $J_{x/y} \subsetneq J_{x'/y'}$.
\end{theorem}

\begin{proof}
All equality and subset relationships (i.e. arcs between notions of
justifiability) depicted in Figure~\ref{fig:hierarchy} are underpinned
by the theorems in section \ref{sec:eq}. Two notions of
justifiability are not subsets of each other iff they are not equal
and not connected by an arc in Figure~\ref{fig:hierarchy}. 
These findings are underpinned by the counter-examples of section
\ref{sec:ex}.
\end{proof}

%

By definition, Prakken and Sartor's semantics~\cite{PS97:Argument},
if we disregard priorities, amounts to $\na{d}/\na{su}$-justifiability.

Similarly, Dung's grounded argumentation semantics~%
\cite{Dun93:ArgumentExplicit} is exactly $\na{a}/\na{u}$-justifiability;
and if we treat explicitly negated literals as new atoms, we can
apply the least fixpoint argumentation semantics for normal logic
programs~\cite{Dun95:Argument,BDKT97:Argument} to extended logic programs, 
which is then, by definition, $\na{u}/\na{u}$-justifiability.

Note that these latter semantics use a slightly different notation to
ours: arguments are sets of assumptions (i.e.~default literals), and a
conclusion of an argument is a literal that can be derived from these
assumptions. This approach can be translated to ours by taking as
arguments all those derivations of a conclusion from an argument. 
Then the definitions of the notions of attack and the fixpoint
semantics coincide. See also the discussion in \cite{PS97:Argument}.

As corollaries to Theorem~\ref{thm:hierarchy} we obtain relationships of
these semantics to the other notions of justifiability.

\begin{corollary}
  Let $J_{Dung}$ be the set of justified arguments according to 
  Dung's grounded argumentation semantics~\cite{Dun93:ArgumentExplicit}.
  Then $J_{Dung}=J_{\na{a}/\na{su}}=J_{\na{a}/\na{u}}=J_{\na{a}/\na{a}}=J_{\na{a}/\na{d}}=J_{\na{a}/\na{sa}}$ and
  $J_{Dung} \subsetneq J_{x/y}$ for all notions of attack $x \not= \na{a}$ and $y$.
  Thus, in Dung's semantics, it does not matter which notion of attack, \na{su,u,a,d,sa}, 
  is used as a defence, and Dung's semantics is more sceptical than
  the others.
\end{corollary}

\begin{corollary}
  Let $J_{PS}$ be the set of justified arguments according to Prakken
  and Sartor's argumentation semantics~\cite{PS97:Argument}, where all
  arguments have the same priority.  Then
  $J_{PS}=J_{\na{d}/\na{su}}=J_{\na{d}/\na{u}}=J_{\na{d}/\na{a}}=J_{\na{d}/\na{d}}=J_{\na{d}/\na{sa}}$,
  $J_{PS} \subsetneq J_{x/y}$ for all notions of attack $x \not\in
  \{\na{a},\na{d}\}$ and $y$, and $J_{PS} \supsetneq J_{\na{a}/y}$ for
  all notions of attack $y$.  Thus, in Prakken and Sartor's semantics,
  it does not matter which notion of attack, \na{su,u,a,d,sa}, is used
  as a defence, and $J_{PS}$ is more credulous than Dung's semantics, but
  more sceptical than all the others.
\end{corollary}

\begin{corollary}
  Let $J_{WFS}$ be the set of justified argument according to the
  well-founded argumentation semantics for normal logic programs
  \cite{Dun95:Argument,BDKT97:Argument}, where an explicitly negated
  atom $\neg L$ is treated as unrelated to the positive atom $L$. Then
  $J_{WFS} = J_{\na{u}/\na{u}} = J_{\na{u}/\na{su}}$, $J_{WFS}
  \supsetneq J_{\na{d}/y} \supsetneq J_{\na{a}/y}$, $J_{WFS}
  \subsetneq J_{\na{su}/y}$, and $J_{WFS} \subsetneq J_{\na{u}/\na{a}}
  = J_{\na{u}/\na{d}} = J_{\na{u}/\na{sa}}$, for all notions of attack
  $y$. Thus, in contrast to Dung's and Prakken and Sartor's semantics,
  for WFS it makes a difference whether rebuts are permitted in the
  defence (\na{a,d,sa}) or not (\na{u,su}).
\end{corollary}

\begin{remark}\hspace{0cm}
  1. The notions of $\na{a}/x$-, $\na{d}/x$- and $\na{sa}/x$-justifiability are 
  particularly sceptical in that even a fact $p$ may not be justified, if there
  is a rule $\neg p \gets B$ (where $not~p \not\in B$) that is not
  $x$-attacked. On the other hand this is useful in terms of
  avoiding inconsistency.

\noindent
2.  $sx/y$-justifiability is particularly credulous,
    because it does not take into account non-strong attacks, so e.g.\
    the program $\{ p \gets not~q, q \gets not~p \}$ has the justified
    arguments $[p \gets not~q]$ and $[q \gets not~p]$.

\end{remark}

\begin{remark}
  One might ask whether any of the semantics in Figure~\ref{fig:hierarchy}
  are equivalent for {\em non-contradictory} programs, i.e.\ programs for
  which there is no literal $L$ such that there exist justified
  arguments for both $L$ and $\neg L$. 
  The answer to this question is no: all the examples in Section \ref{sec:ex} distinguishing
  different notions of justifiability involve only non-contradictory 
  programs.

  In particular, even for non-contradictory programs,
  Dung's and Prakken and Sartor's semantics differ, and
  both differ from $\na{u}/\na{a}$-justifiability, which will be shown equivalent to
  the paraconsistent well-founded semantics WFSX$_p$~\cite{Dam96:Thesis,PA92:WFSX,AP96:WFSX} in
  Section~\ref{sec:wfsxarg}.
\end{remark}

\section{Properties of Argumentation Semantics}
\label{sec:properties}

We will now state some important properties which a semantics for
extended logic programs may have, and examine for which of the
argumentation semantics these properties hold.

\subsection{The coherence principle}
\label{sec:coherence}

The coherence principle for extended logic
programming~\cite{AP96:WFSX} states that ``explicit negation implies
implicit negation''.  If the intended meaning of $not~L$ is ``if there
is no evidence for $L$, assume that $L$ is false'', and the intended
meaning of $\neg L$ is ``there is evidence for the falsity of $L$'',
then the coherence principle states that explicit evidence is
preferred over assumption of the lack of evidence.  Formally, this can
be stated as: if $\neg L$ is in the semantics, then $not~L$ is also in
the semantics. In an argumentation semantics, we have not defined what
it means for a default literal to be ``in the semantics''. This can
easily be remedied, though, and for convenience we introduce the
following transformation.\footnote{The purpose of the transformation
could be equally achieved by defining that $not~L$ is $x/y$-justified
if all arguments for $L$ are overruled.}

\begin{definition}
  Let $P$ be an extended logic program, and $x$ and $y$ notions of attack, 
  and let $L$ be an objective literal.
  Then $L$ is {\em $x/y$-justified} if there exists
  a $x/y$-justified argument for $L$. 

  Let $nL$ be a fresh atom, and $P' = P \cup \{ nL \gets not~L \}$.
  Then $not~L$ is {\em $x/y$-justified} if $[ nL \gets not~L ]$ is
  a $x/y$-justified argument associated with $P'$.
\end{definition}
Note that because $nL$ is fresh, then either $J_{x/y}(P') = J_{x/y}(P)$ or 
$J_{x/y}(P') = J_{x/y}(P) \cup \{ [ nL \gets not~L ] \}$.

\begin{definition}
  A least fixpoint semantics $J_{x/y}$  
  {\em satisfies the coherence principle} if 
  for every objective literal $L$,
  if $\neg L$ is $x/y$-justified, then $not~L$ is 
  $x/y$-justified. 
\end{definition}

The following result states that a least fixpoint semantics satisfies
the coherence principle exactly in those cases where we allow any
attack for the defence.  
Informally, this is because the only way of attacking a default
literal $not~L$ is by undercut, i.e.\ an argument for $L$, and in
general, such an argument can only be attacked by an argument for
$\neg L$ by a rebut.

\begin{theorem}
  Let $x,y \in \{ \na{a}, \na{u}, \na{d}, \na{su}, \na{sa} \}$.
  Then $J_{x/y}$ satisfies the coherence principle
  iff $J_{x/y} = J_{x/\na{a}}$.
\end{theorem}

\begin{proof}
  \begin{itemize}
  \item For the ``only if'' direction, we show that for those notions
    of justifiability $x/y \not= x/\na{a}$, the coherence principle does 
    not hold. 

  \begin{itemize}
  \item Consider the program $P$:
        
    $\begin{array}{rcl}
      p & \gets & not~q \\
      q & \gets & not~r \\
      r & \gets & not~s \\
      s & \gets & not~p \\
      \neg p \\
    \end{array}$
      
    Then $J_{\na{u}/\na{u}}(P') = J_{\na{su}/\na{u}}(P') =
    J_{\na{su}/\na{su}}(P') = \{ [\neg p] \}$, 
    where $P' = P \cup \{ np \gets not~p \}$.
    In these cases, the coherence principle is not satisfied, because
    $\neg p$ is justified, but $not~p$ is not justified.
  \item Now consider the program $Q$:
    
    $\begin{array}{rcl}
      p & \gets & not~\neg p \\
      \neg p & \gets & not~p \\
    \end{array}$

    Then $J_{\na{su}/\na{sa}}(Q') = J_{\na{sa}/\na{sa}}(Q') =
    \{ [p \gets not~\neg p], [\neg p \gets not~p] \}$,
    where $Q' = Q \cup \{ np \gets not~p \}$.
    Again, the coherence principle is not satisfied, because
    $\neg p$ is justified, but $not~p$ is not justified.
    \\[1ex]

  \end{itemize}

   \item For the ``if'' direction, let $x$ be any notion of attack.
    Let $P$ be an extended logic program, and 
    $\neg L$ a $x/\na{a}$-justified literal,
    i.e.\ there is an argument $A = [ \neg L \gets Body, \ldots ]$
    and an ordinal $\alpha$ s.t.~$A \in J_{x/\na{a}}^\alpha$.

    Let $A' = [ nL \gets not~L ]$, and $(B,A') \in x$.
    Because $nL$ is fresh, the only possible attack on $A'$ is 
    a strong undercut, i.e.\ $L$ is a conclusion of $B$.
    Then $A$ attacks $B$, and so $[ nL \gets not~L ] \in
    J_{x/\na{a}}^{\alpha+1}$.
    
  \end{itemize}
\end{proof}

\subsection{Consistency}
\label{sec:consistency}

Consistency is an important property of a logical system. It states
that the system does not support contradictory conclusions. In
classical logic ``ex falso quodlibet'', i.e.\ if both $A$ and $\neg A$
hold, then any formula holds. 
In paraconsistent systems~\cite{DP98:Paraconsistent}, this property
does not hold, thus allowing both $A$ and $\neg A$ to hold for a
particular formula $A$, while not supporting any other contradictions.

A set of arguments is \emph{consistent}, or 
\emph{conflict-free}~\cite{PS97:Argument,Dun95:Argument},
if it does not contain two arguments
such that one attacks the other. There are several notions of
consistency, depending on which notion of attack is considered
undesirable.

\begin{definition}
  Let $x$ be a notion of attack, and $P$ an extended logic program. 
  Then a set of arguments associated with $P$ is called 
  \emph{$x$-consistent} if it does not contain arguments $A$ and $B$ such
  that $(A,B) \in x_P$.
\end{definition}

The argumentation semantics of an extended logic program need not
necessarily be consistent; because of explicit negation, there exist
contradictory programs such as $\{ p, \neg p \}$, for which there
exist sensible, but inconsistent arguments ($[p]$ and $[\neg p]$ in
this case).

A general result identifies cases in which the set of justified arguments
for a program is consistent. It states that if we allow the attack to
be at least as strong as the defence, i.e.\ if we are
{\em sceptical}, then the set of justified arguments is
consistent. 

\begin{theorem}
  Let $x$ and $y$ be notions of attack such that $x \!\supseteq\! y$, 
  and let $P$ be an extended logic program.
  Then the set of $x/y$-justified arguments is $x$-consistent.
\end{theorem}

\begin{proof}
  We show that $J_{x/y}^\alpha$ is $x$-consistent for all 
  ordinals $\alpha$, by transfinite induction on $\alpha$.

  \noindent {\it Base case} $\alpha=0$: Trivial.

  \noindent {\it Successor ordinal} $\alpha \leadsto \alpha+1$:
  Assume $A,B \in J_{x/y}^{\alpha+1}$ and $(A,B) \in x$.
  Then there exists $C \in J_{x/y}^\alpha$ such that
  $(C,A) \in y \subseteq x$. Then by induction hypothesis, 
  because $C \in J_{x/y}^\alpha$, then $A \not\in J_{x/y}^\alpha$.
  Because $A \in J_{x/y}^{\alpha+1}$, there exists
  $D \in J_{x/y}^\alpha$ such that $(D,C) \in y \subseteq x$.
  This contradicts the induction hypothesis, so we have to retract
  the assumption and conclude that $J_{x/y}^{\alpha+1}$ is
  $x$-consistent.

  \noindent {\it Limit ordinal $\lambda$:}
  Assume $A, B \in J_{x/y}^\lambda$ and $(A,B) \in x$.
  Then there exist $\alpha, \beta < \lambda$ s.t.~$A \in J_{x/y}^\alpha$ and
  $B \in J_{x/y}^\beta$. W.l.o.g. assume that $\alpha \leq \beta$.
  Then because $J_{x/y}^\alpha \subseteq J_{x/y}^\beta$, we have
  $A \in J_{x/y}^\beta$, contradicting the induction hypothesis that 
  $J_{x/y}^\beta$ is $x$-consistent.
\end{proof}

The following example shows that, in general, the set of justified
arguments may well be inconsistent.

\begin{example} 
Consider the following program:
\begin{center}
    $\begin{array}{rcl}
      q & \gets & not~p \\
      p \\
      \neg p
    \end{array}$
\end{center}
Then $J_{\na{u}/\na{a}} = \{ [q \gets not~p], [p], [\neg p] \}$,
  and $[p]$ and $[\neg p]$ rebut each other, and
  $[p]$ strongly undercuts $[q \gets not~p]$.
\end{example}

\section{Argumentation Semantics and WFSX}
\label{sec:wfsxarg}

In this section we will prove that the argumentation semantics
$J_{\na{u}/\na{a}}$ is equivalent to the paraconsistent well-founded semantics
with explicit negation WFSX$_p$~\cite{Dam96:Thesis,AP96:WFSX}.
First, we summarise the definition of WFSX$_p$.

\subsection{Well-founded semantics with explicit negation}
\label{sec:wfsx}

We recollect the definition of the paraconsistent well-founded semantics for 
extended logic programs, WFSX$_p$. We use the definition of~\cite{ADP95:LPsystem},
because it is closer to our definition of argumentation semantics than
the original definition of~\cite{PA92:WFSX}.

\begin{definition} 
  The set of all objective literals of a program $P$ is 
  called the {\em Herbrand base} of $P$ and denoted by $\herbrand(P)$.
  A {\em paraconsistent interpretation} of a program $P$ is a set
  $T \cup not~F$ where $T$ and $F$ are subsets of $\herbrand(P)$.
  An {\em interpretation} is a paraconsistent interpretation where the
  sets $T$ and $F$ 
  are disjoint. An interpretation is called {\em two-valued} if
  $T \cup F = \herbrand(P)$.
\end{definition}

\begin{definition}
  Let $P$ be an extended logic program, $I$ an interpretation, and let
  $P'$ (resp.~$I'$) be obtained from $P$ (resp.~$I$) by replacing
  every literal $\neg A$ by a new atom, say $\neg\_A$.  The
  GL-transformation $\frac{P'}{I'}$ is the program obtained from $P'$
  by removing all rules containing a default literal $not~A$ such that
  $A \in I'$, and then removing all remaining default literals from
  $P'$, obtaining a definite program $P''$.  Let $J$ be the least
  model of $P''$, i.e. $J$ is the least fixpoint of $T_{P''}(I) :=
  \{A~|~~\exists A\gets B_1,\dots,B_n \in P'' \mbox{ s.t. } B_i \in
  I\}$. Then $\Gamma_P I$ is obtained from $J$ by replacing the introduced
  atoms $\neg\_A$ by $\neg A$.
\end{definition}

\begin{definition}
  The {\em semi-normal} version of a program $P$ is the program $P_s$
  obtained from $P$ by replacing every rule $L \gets Body$ in $P$ by
  the rule $L \gets not~\neg L, Body$.  If the program $P$ is clear
  from the context, we write $\Gamma I$ for $\Gamma_P I$ and $\Gamma_s
  I$ for $\Gamma_{P_s} I$.
\end{definition}

Note that the set $\Gamma_P I$ is just a set of literals; we will now
use it to define the semantics of $P$ as a (paraconsistent) interpretation.

\begin{definition}
  Let $P$ be a program whose least fixpoint of $\Gamma \Gamma_s$ is $T$.
  Then the {\em paraconsistent well-founded model of $P$} is
  the paraconsistent interpretation
  $WFM_p(P) = T \cup not~(\herbrand(P) - \Gamma_s T)$.
  If $WFM_p(P)$ is an interpretation, then
  $P$ is called {\em non-contradictory},
  and $WFM_p(P)$ is the {\em well-founded model of $P$},
  denoted by $WFM(P)$.
\end{definition}
The paraconsistent well-founded model can be defined iteratively by
the transfinite sequence $\{I_\alpha\}$:

  \begin{tabular}{llll}
    $I_0$ & := & $\emptyset$ \\
    $I_{\alpha+1}$ & := & $\Gamma \Gamma_s I_\alpha$ &
    for successor ordinal $\alpha+1$ \\
    $I_\lambda$ & := & $\bigcup_{\alpha < \lambda} I_\alpha$ & 
    for limit ordinal $\lambda$ \\
  \end{tabular}

\noindent
There exists a smallest ordinal $\lambda_0$ such that $I_{\lambda_0}$
is the least fixpoint of $\Gamma \Gamma_s$, and \linebreak
$WFM_p(P) := I_{\lambda_0} \cup not~(\herbrand(P) - \Gamma_s I_{\lambda_0})$.

\subsection{Equivalence of argumentation semantics and WFSX$_p$}

In this section, we will show that the argumentation semantics
$J_{\na{u}/\na{a}}$ and the well-founded model coincide.  That is, the
conclusions of justified arguments are exactly the objective literals
which are true in the well-founded model; and those objective literals
all of whose arguments are overruled are exactly the literals which
are false in the well-founded model.  The result holds also for
contradictory programs under the {\em paraconsistent} well-founded
semantics.  This is important, because it shows that contradictions in
the argumentation semantics are precisely the contradictions under the
well-founded semantics, and allows the application of contradiction
removal (or avoidance) methods to the argumentation semantics
\cite{sch97c}. For non-contradictory programs, the
well-founded semantics coincides with the paraconsistent well-founded
semantics~\cite{AP96:WFSX,Dam96:Thesis}; consequently, we obtain as a corollary that
argumentation semantics and well-founded semantics coincide for
non-contradictory programs.

Before we come to the main theorem, we need
the following Lemma, which shows a precise connection between arguments and
consequences of a program $\frac{P}{I}$.
\begin{lemma}
  \label{lemma:model-arg}
  Let $I$ be a two-valued interpretation.
  \begin{enumerate}
  \item \label{lemma:model-arg:u}
    $L \in \Gamma(I)$ iff $\exists$ argument $A$ with conclusion $L$
    such that $\assm(A) \subseteq I$.
  \item \label{lemma:model-arg:a}
    $L \in \Gamma_s(I)$ iff $\exists$ argument $A$ with conclusion $L$
    such that $\assm(A) \subseteq I$ and
    $\neg\conc(A) \cap I = \emptyset$.
  \item \label{lemma:model-arg:u-neg}
    $L \not\in \Gamma(I)$ iff $\forall$ arguments $A$ with conclusion $L$,
    $\assm(A) \cap I \not\subseteq \emptyset$.
  \item \label{lemma:model-arg:a-neg}
    $L \not\in \Gamma_s(I)$ iff $\forall$ arguments $A$ with 
    conclusion $L$, $\assm(A) \cap I \not\subseteq \emptyset$
    or $\neg\conc(A) \cap I \not= \emptyset$.
  \end{enumerate}
\end{lemma}
\begin{proof} 
  See \ref{appendix}.
\end{proof}

In order to compare the argumentation semantics with the well-founded
semantics, we extend the definition $\conc(A)$ of the conclusions of a
single argument $A$ to work on a set of arguments $\mathcal{A}$. 
The extended definition $\conc(\mathcal{A})$ includes all positive and
negative conclusions of arguments in $\mathcal{A}$; i.e. those
literals $L \in \conc(\mathcal{A})$, as well as the default literals
$not~L$ where all arguments for $L$ are overruled by some argument
$A\in\mathcal{A}$. 
We will use this definition of $\conc$ for the set
of justified arguments $J_{\na{u/a}}$ to compare the ``argumentation
model'' $\conc(J_{\na{u/a}})$ to $WFM_p(P)$, the well-founded model.

\begin{definition}
Let $\mathcal{A}$ be a set of arguments. Then 
$$\conc(\mathcal{A}) = \bigcup_{A\in\mathcal{A}} \conc(A) \cup \{ not~L~~|~ \mbox{all arguments for } L \mbox{ are overruled by an argument } A\in\mathcal{A}\}$$
\end{definition}

With the above definition, we can formulate the main theorem that
$\na{u/a}$-justified arguments coincide with the well-founded
semantics.

\begin{theorem}
  \label{thm:wfsxarg}
  Let $P$ be an extended logic program.
  Then $WFM_p(P) = \conc(J_{\na{u/a}})$.
\end{theorem}

\begin{proof}
  First, note that $A$ undercuts $B$ iff 
  $\exists~L$ s.t.\ $L \in \conc(A)$ and $not~L \in \assm(B)$;
  and $A$ rebuts $B$ iff $\exists~L \in \conc(A) \cap \neg\conc(B)$.\\[2ex]
  We show that for all ordinals $\alpha$, $I_\alpha =
  \conc(J^\alpha_{\na{u/a}})$, by transfinite induction on
  $\alpha$. The proof proceeds in two stages. First, we show that all
  objective literals $L$ in $WFM_p(P)$ are conclusions of
  $\na{u/a}$-justified arguments and second, that for all default
  negated literals $not~L$ in $WFM_p(P)$, all arguments for $L$ are
  overruled. \\[0.5ex]
  \textit{Base case $\alpha=0$:}
  $I_\alpha = \emptyset = \conc(J^\alpha_{\na{u/a}})$
  \\[0.5ex]
  \textit{Successor ordinal $\alpha \leadsto \alpha+1$}:\\[0.5ex]
  $L \in I_{\alpha+1}$ \\ 
  \mycenter{iff (Def.\ of $I_{\alpha+1}$)} \\
  $L \in \Gamma \Gamma_s I_\alpha$  \\
  \mycenter{iff (Lemma~\ref{lemma:model-arg}(\ref{lemma:model-arg:u}))} \\
  $\exists$~argument $A$ for $L$ such that
  $\assm(A) \subseteq \Gamma_s I_\alpha$ \\
  \mycenter{iff (Def.\ of $\subseteq$, and $\Gamma_s I_\alpha$ 
    is two-valued)} \\
  $\exists$~argument $A$ for $L$ such that
  $\forall~not~L \in \assm(A), L \not\in \Gamma_s I_\alpha$ \\ 
  \mycenter{iff (Lemma~\ref{lemma:model-arg}(\ref{lemma:model-arg:a-neg}))} \\
  $\exists$~argument $A$ for $L$ such that $\forall~not~L \in \assm(A)$, 
  for any argument $B$ for $L$,
  ( $\exists~not~L' \in \assm(B) ~s.t.\ L' \in I_\alpha$
    or $\exists~L'' \in \conc(B) ~s.t.\ \neg L'' \in I_\alpha$ ) \\
  \mycenter{iff (Induction hypothesis)} \\
  $\exists$~argument $A$ for $L$ such that $\forall~not~L \in \assm(A)$, 
  for any argument $B$ for $L$,
  ( $\exists~not~L' \in \assm(B) ~s.t.\ \exists$
    argument $C \in J^\alpha_{\na{u/a}}$ for $L'$,
    or $\exists~L'' \in \conc(B) ~s.t.\ \exists$~
    argument $C \in J_{\na{u/a}}^\alpha$ for $\neg L''$) \\
  \mycenter{iff (Def.\ of undercut and rebut)} \\
  $\exists$~argument $A$ for $L$ such that for any undercut $B$ to $A$,
  ( $\exists$~argument $C \in J_{\na{u/a}}^\alpha$ s.t.\
    $C$ undercuts $B$,
    or $\exists$~argument $C \in J_{\na{u/a}}^\alpha$ s.t.\
    $C$ rebuts $B$) \\
  \mycenter{iff} \\
  $\exists$~argument $A$ for $L$ such that for any undercut $B$ to $A$,
  $\exists$~argument $C \in J_{\na{u/a}}^\alpha$ s.t.\ $C$ attacks $B$ \\
  \mycenter{iff (Def.\ of $J_{\na{u/a}}^{\alpha+1}$)} \\
  $\exists$~argument $A \in J_{\na{u/a}}^{\alpha+1}$ for $L$ \\
  \mycenter{iff  (Def.\ of $\conc$)}\\
  $L \in \conc(J_{\na{u/a}}^{\alpha+1})$\\[0.5ex]
  \textit{Limit ordinal $\lambda$:}\\
  $I_\lambda = \bigcup_{\alpha < \lambda} I_\alpha$ and
  $J_{\na{u/a}}^\lambda = \bigcup_{\alpha < \lambda} J_{\na{u/a}}^\alpha$,
  so by induction hypothesis ($I_\alpha = \conc(J_{\na{u/a}}^\alpha)$ for all $\alpha < \lambda$),
  $I_\lambda = \conc(J_{\na{u/a}}^\lambda)$.\\[2ex]
  Next we will show that a literal $not~L$ is in the well-founded semantics
  iff every argument for $L$ is overruled, i.e.
  $not~L \in WFM_p(P)$ implies
  $not~L \in \conc(J_{\na{u/a}})$.\\[0.5ex]
  $not~L \in WFM_p(P)$ \\
  \mycenter{iff (Def.\ of $WFM_p(P)$)} \\
  $L \not\in \Gamma_s I_\lambda$ \\
  \mycenter{iff (Lemma~\ref{lemma:model-arg}(\ref{lemma:model-arg:a-neg})} \\
  for all arguments $A$ for $L$,
  ( $\exists~not~L' \in \assm(A) ~s.t.\ L' \in I_\lambda$, or
    $\exists~L'' \in \conc(A) ~s.t.\ \neg L'' \in I_\lambda$ ) \\
  \mycenter{iff ($I_\lambda=\conc(J_{\na{u/a}}^\lambda)$)} \\
  for all arguments $A$ for $L$,
  ( $\exists ~not~L' \in \assm(A) ~s.t.\ 
    \exists$~argument $B \in J_{\na{u/a}}^\lambda$ for $L'$, or
    $\exists~L'' \in \conc(A) ~s.t.\ 
    \exists$~argument $B \in J_{\na{u/a}}^\lambda$ for $\neg L''$ ) \\
  \mycenter{iff (Def.\ of undercut and rebut)} \\
  for all arguments $A$ for $L$,
  ( $\exists$~argument $B \in J_{\na{u/a}}^\lambda$ s.t.\ $B$ undercuts $A$, or
    $\exists$~argument $B \in J_{\na{u/a}}^\lambda$ s.t.\ $B$ rebuts $A$ ) \\
  \mycenter{iff} \\
  every argument for $L$ is attacked by a justified argument in $J_{\na{u/a}}^\lambda$ \\
  \mycenter{iff (Def.\ of overruled)} \\
  every argument for $L$ is overruled \\
  \mycenter{iff (Def.\ of $\conc(J_{\na{u/a}})$)} \\
  $not~L \in \conc(J_{\na{u/a}})$
\end{proof}

\begin{corollary}
  Let $P$ be a non-contradictory program. Then $WFM(P) = \conc(J_{\na{u/a}})$.
\end{corollary}

\begin{remark}
  In a similar way, one can show that the $\Gamma$ operator corresponds
  to undercuts, while the $\Gamma_s$ operator corresponds to attacks, and so
  the least fixpoints of $\Gamma\Gamma$, $\Gamma_s\Gamma$, and
  $\Gamma_s\Gamma_s$ correspond to $J_{\na{u}/\na{u}}$, $J_{\na{a}/\na{u}}$, and $J_{\na{a}/\na{a}}$,
  respectively. 
  In~\cite{ADP95:LPsystem}, the least fixpoints of these operators are
  shown to be ordered as 
  $lfp(\Gamma_s\Gamma) \subseteq lfp(\Gamma_s\Gamma_s) \subseteq
   lfp(\Gamma\Gamma_s)$, and 
  $lfp(\Gamma_s\Gamma) \subseteq lfp(\Gamma\Gamma) \subseteq 
   lfp(\Gamma\Gamma_s)$. 
  Because $J_{\na{a}/\na{u}} = J_{\na{a}/\na{a}} \subseteq J_{\na{u}/\na{u}} \subseteq J_{\na{u}/\na{a}}$ by 
  Theorem~\ref{thm:hierarchy}, we can strengthen this statement
  to $lfp(\Gamma_s\Gamma) = lfp(\Gamma_s\Gamma_s) \subseteq
  lfp(\Gamma\Gamma) \subseteq lfp(\Gamma\Gamma_s)$.
\end{remark}

The following corollary summarises the results so far.

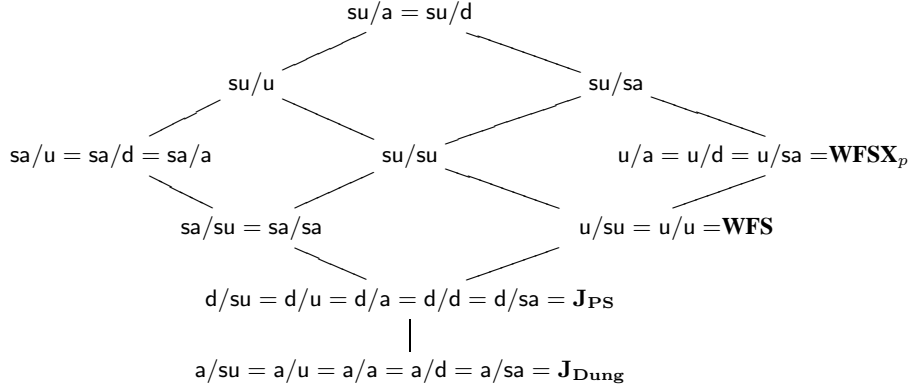
\begin{figure}[htbp]
  \begin{center}
    $\xymatrix@C=-2em@R=3ex{
      && \na{su}/\na{a} = \na{su}/\na{d} \\
      & \na{su}/\na{u} \ar@{-}[ru] && \na{su}/\na{sa} \ar@{-}[lu] \\
      \na{sa}/\na{u} = \na{sa}/\na{d} = \na{sa}/\na{a}
      \ar@{-}[ru] && 
      \na{su}/\na{su} \ar@{-}[lu] \ar@{-}[ru] &&
      {\makebox[10em]{\hspace{-5em}$\na{u}/\na{a} = \na{u}/\na{d} = \na{u}/\na{sa} = $\textbf{WFSX$_p$}}}
      \ar@{-}[lu] \\
      & \na{sa}/\na{su} = \na{sa}/\na{sa} \ar@{-}[lu] \ar@{-}[ru] && 
      {\makebox[10em]{\hspace{5em}$\na{u}/\na{su} = \na{u}/\na{u} = $\textbf{WFS}}} 
      \ar@{-}[lu] \ar@{-}[ru] \\
      && {\makebox[10em]{$\na{d}/\na{su} = \na{d}/\na{u} = \na{d}/\na{a} = \na{d}/\na{d} = \na{d}/\na{sa} = 
          \mathbf{J_{PS}}$}}
      \ar@{-}[lu] \ar@{-}[ru] \\
      && {\makebox[10em]{$\na{a}/\na{su} = \na{a}/\na{u} = \na{a}/\na{a} = \na{a}/\na{d} = \na{a}/\na{sa} = 
          \mathbf{J_{Dung}}$}}
      \ar@{-}[u] \\
      }$
    \caption{Hierarchy of Notions of Justifiability and Existing Semantics}
    \label{fig:compare-hierarchy}
  \end{center}
\end{figure}

\begin{corollary}
  The least fixpoint argumentation semantics of 
  Dung~\cite{Dun93:ArgumentExplicit}, denoted $\mathbf{J_{Dung}}$,
  of Prakken and Sartor~\cite{PS97:Argument}, denoted $\mathbf{J_{PS}}$, 
  and the well-founded semantics
  for normal logic programs \textbf{WFS}~\cite{BDKT97:Argument,GRS91:WFS} 
  and for logic programs with explicit negation 
  \textbf{WFSX$_p$}~\cite{PA92:WFSX,AP96:WFSX} are related to 
  the other least fixpoint argumentation semantics as illustrated in
  Figure~\ref{fig:compare-hierarchy}.
\end{corollary}

\section{Proof Theory}
\label{sec:proof-theory}

One of the benefits of relating the argumentation semantics $J_{\na{u}/\na{a}}$ to
WFSX$_p$ is the existence of an efficient top-down proof procedure for 
WFSX$_p$~\cite{ADP95:LPsystem},
which we can use to compute justified arguments in $J_{\na{u}/\na{a}}$.
On the other hand, {\em dialectical} proof theories, based on dialogue trees,
have been defined for a variety of argumentation semantics
\cite{SCG94:Dialectics,PS97:Argument,JV99:DialecticSemantics,KT99:ComputingArgumentation}.
In this section we present a sound and complete dialectical proof theory for
the least fixpoint argumentation semantics $J_{x/y}$ for any
notions of attack $x$ and $y$.

\subsection{Dialogue trees}

We adapt the dialectical proof theory of~\cite{PS97:Argument} to
develop a general sound and complete proof theory for 
$x/y$-justified arguments.

\begin{definition}
\label{def:dialog}
  Let $P$ be an extended logic program.
  An {\em $x/y$-dialogue} is a finite nonempty sequence of 
  moves $move_i = (\mathit{Player}_i,Arg_i) (i > 0)$, such that
  $Player_i \in \{P,O\}$, $Arg_i \in \mathit{Args}_P$, and 
  \begin{enumerate}
  \item $\mathit{Player}_i = P$ iff $i$ is odd; and 
    $\mathit{Player}_i = O$ iff $i$ is even.
  \item If $\mathit{Player}_i = \mathit{Player}_j = P$ and $i \not= j$,
    then $Arg_i \not= Arg_j$.
  \item \label{item:minimal} If $\mathit{Player}_i = P$ and $i > 1$, then
    $Arg_i$ is a minimal argument such that $(Arg_i,Arg_{i-1}) \in y$.
  \item If $\mathit{Player}_i = O$, then $(Arg_i,Arg_{i-1}) \in x$.
  \end{enumerate}
\end{definition}

The first condition states that the players $P$ (Proponent) and
$O$ (Opponent) take turns, and $P$ starts.
The second condition prevents the proponent from repeating a move.
The third and fourth conditions state that both players have to 
attack the other player's last move, where the opponent is allowed
to use the notion of attack $x$, while the proponent may use $y$ to 
defend its arguments.
Note that the minimality condition in~\ref{item:minimal} is redundant,
because {\em all} arguments in $\mathit{Args}_P$ are required 
to be minimal by Definition~\ref{def:argument}.
We have explicitly repeated this condition, because it is important
in that it prevents the proponent from repeating an argument by adding
irrelevant rules to it.

\begin{definition}
\label{def:dialogtree}
  An {\em $x/y$-dialogue tree} is a tree of moves such that
  every branch is a $x/y$-dialogue, and for all moves $move_i = (P, Arg_i)$,
  the children of $move_i$ are all those moves $(O,Arg_j)$ such that
  $(Arg_j,Arg_i) \in x$.\\
  The {\em height} of a dialogue tree is $0$ if it consists only of
  the root, and otherwise $height(t) = sup \{ height(t_i) \} + 1$ where 
  $t_i$ are the trees rooted at the grandchildren of $t$.
\end{definition}

\begin{example}
  Consider the following program:
\begin{center}
    $\begin{array}{rcl}
     p & \gets & q, not~r \\
     q & \gets & not~s \\
     \neg q & \gets & u\\
     r & \gets & not~t \\
     s & \gets & not~t \\
     t & \gets & not~w \\
     u & \gets & not~v \\
     v & \gets & not~r \\
     \neg v & \gets & not~t
    \end{array}$
\end{center}
A $\na{a}/\na{u}$-dialogue tree rooted at the argument
$[p \gets q, not~r; q \gets not~s]$ is given by Figure~\ref{fig:tree}.
Each node is marked with $P$ for proponent or $O$ for opponent,
and an edge $\xymatrix{A \ar[r]^x & B}$ denotes that $A$ attacks $B$
with the notion of attack $x$, i.e.~$(A,B) \in x$.
\begin{figure}[htbp]
  $$\xymatrix{
    & P:[p \gets q, not~r; q \gets not~s] \\
    O:[r \gets not~t] \ar[ru]^{\na{u}} &
    O:[\neg q \gets u; u \gets not~v] \ar[u]^{\na{r}} &
    O:[s \gets not~t] \ar[lu]_{\na{u}} \\
    P:[t \gets not~w] \ar[u]^{\na{u}} &
    P:[v \gets not~r] \ar[u]^{\na{u}} &
    P:[t \gets not~w] \ar[u]^{\na{u}} \\
    & O:[r \gets not~t] \ar[u]^{\na{u}} &
    O:[\neg v \gets not~t] \ar[lu]_{\na{r}} \\
    & P:[t \gets not~w] \ar[u]^{\na{u}} &
    P:[t \gets not~w] \ar[u]^{\na{u}} \\
  }$$
    \caption{An $\na{a}/\na{u}$-dialogue tree}
    \label{fig:tree}
\end{figure}
\end{example}
Note that although dialogues are required to be finite, dialogue trees
may be infinitely branching. Therefore dialogue trees need not be
finite, nor need their height be finite.

\begin{example} 
Consider the following program $P$
\footnote{Note that by definition, programs are not allowed to contain
variables. Here, $X$ denotes a variable, and $P$ is an abbreviation
for the (infinite) program obtained by substituting the terms $s^n(0)$
for the variable $X$, in all the rules.}:
\begin{center}
    $\begin{array}{rcl}
      p(0) \\
      p(s(X)) & \gets & not~q(X) \\
      q(X) & \gets & not~p(X) \\
      r & \gets & q(X) \\
      s & \gets & not~r \\
    \end{array}$
\end{center}
For each $n \in \mathbb{N}$, there is exactly one minimal argument $A_n$
with conclusion $p(s^n(0))$, namely $[p(0)]$ for $n=0$, and
$[p(s^n(0)) \gets not~q(s^{n-1}(0))]$ for $n > 0$.
Similarly, there is exactly one minimal argument $B_n$ with conclusion
$q(s^n(0))$, namely $[q(s^n(0)) \gets not~p(s^n(0))]$.\\
Therefore, a $\na{u}/\na{u}$-dialogue tree rooted at $A_{n+1}$ consists 
of just one dialogue $T_{n+1}$ of the form 
$( (P, A_{n+1}), (O, B_{n}), T_n )$. A $\na{u}/\na{u}$-dialogue tree
rooted at $A_0$ consists only of the root, because there are no
undercuts to $A_0$. Thus, the height of the dialogue tree $T_n$ is $n$.\\
Now consider the $\na{u}/\na{u}$-dialogue tree rooted at the argument
$C = [s \gets not~r]$. The argument $C$ is undercut by infinitely many
arguments $D_n = [r \gets q(s^n(0)); q(s^n(0)) \gets not~p(s^n(0))]$;
each $D_n$ is undercut by exactly one argument: $A_n$.
A dialogue in the $\na{u}/\na{u}$-dialogue tree $T_C$ rooted at
argument $C$ is therefore a sequence $( (P, C), (O, B_n), T_n )$. 
Because $\mathit{height}(T_n)=n$, then by Definition~\ref{def:dialogtree}:
$\mathit{height}(T_C) = 
  \mathit{sup} \{ \mathit{height}(T_n) ~|~ n \in \mathbb{N} \} + 1 =
  \omega + 1$.
\end{example}

\begin{definition}
\label{def:windialogtree}
  A player {\em wins an $x/y$-dialogue} iff the other player cannot move.
  A player {\em wins an $x/y$-dialogue tree} iff it wins all branches
  of the tree. An $x/y$-dialogue tree which is won by the proponent
  is called a {\em winning $x/y$-dialogue tree}.
\end{definition}

We show that the proof theory of $x/y$-dialogue trees is sound and
complete for any notions of attack $x$ and $y$.

\begin{theorem}
  \label{thm:sound-complete}
  An argument $A$ is $x/y$-justified iff there exists a $x/y$-dialogue
  tree with $A$ as its root, and won by the proponent.
\end{theorem}

\begin{proof}
  We show by transfinite induction that for all arguments $A$,
  for all ordinals $\alpha$:
  $A \in J^{\alpha}_{x/y}$ if and only if
  there exists a winning $x/y$-dialogue tree of height $\leq \alpha$ 
  for $A$.
  See \ref{appendix} for the detailed proof.
\end{proof}

\section{Related Work}

There has been much work on argument-theoretic semantics for
normal logic programs, i.e. logic programs with default negation~%
\cite{BDKT97:Argument,Dun95:Argument,KT99:ComputingArgumentation}.
Because there is no explicit negation, there is only one form of
attack, the {\em undercut} in our terminology.
An abstract argumentation framework has been defined, which captures
other default reasoning mechanisms besides normal logic programming.
Within this framework, a variety of semantics may be defined,
such as \emph{preferred extensions}; \emph{stable extensions}, which are
equivalent to \emph{stable models}~\cite{GL88:StableModels};
and a least fixpoint semantics based on the acceptability of arguments,
which is equivalent to the \emph{well-founded semantics}~\cite{GRS91:WFS}.
The latter fixpoint semantics forms the basis of our argumentation semantics.
Proof theories and proof procedures for some of these argumentation
semantics have been developed in~\cite{KT99:ComputingArgumentation}.

There has been some work extending this argumentation semantics
to logic programs with explicit negation. Dung~\cite{Dun95:Argument}
adapts the framework of~\cite{Dun93:ArgumentExplicit}, by distinguishing
between {\em ground attacks} and {\em reductio-ad-absurdum-attacks},
in our terminology undercuts and rebuts. Argumentation semantics analogous to
those of normal logic programs are defined, and the stable extension
semantics is shown to be equivalent to the answer set 
semantics~\cite{GL90:ClassicalNegation}, an adaptation of the stable
model semantics to extended logic programs. A least fixpoint semantics
(called {\em grounded} semantics) based on a notion of acceptability 
is defined, and related to the well-founded semantics of~\cite{GRS91:WFS}, 
although only for the case of programs without explicit negation.

Prakken and Sartor~\cite{PS97:Argument} define an argumentation
semantics for extended logic programs similar to that of Dung.  
Their language is more expressive in that it distinguishes between 
{\em strict} rules, which may not be attacked, and {\em defeasible}
rules, which may be attacked.  Furthermore, rules have priorities, and
rebuts are only permitted against a rule of equal or lower priority.
Thus, rebuts are not necessarily symmetric, as in our setting.  Our
language corresponds to Prakken and Sartor's without strict rules, 
and either without priorities,
or, equivalently, if all rules have the same priority.  The semantics
is given as a least fixpoint of an acceptability operator, analogous
to Dung's grounded semantics.  A proof theory, similar to those of
Kakas and Toni~\cite{KT99:ComputingArgumentation} is developed. This
proof theory formed the basis of our general proof theory for
justified arguments.

In \cite{MA98:Argumentation}, an argumentation semantics for extended
logic programs, similar to
Prakken and Sartor's, is proposed; it is influenced by WFSX, and
distinguishes between sceptical and credulous conclusions of an
argument. It also provides a proof theory based on dialogue trees,
similar to Prakken and Sartor's.

Defeasible Logic Programming 
\cite{GS03:DeLP,SCG94:Dialectics,GSC98:Argument}
is a formalism very similar to Prakken and Sartor's, 
based on the first order logic argumentation framework of \cite{SL92:MTDR}. 
It includes logic programming with two kinds of negation,
distinction between strict and defeasible rules, and allowing for
various criteria for comparing arguments. 
Its semantics is given operationally, by proof procedures based on
dialectical trees \cite{GS03:DeLP,SCG94:Dialectics}.
In \cite{DCSS02:DefeasibleTransformation}, the semantics of Defeasible
Logic Programming is related to the well-founded semantics, albeit
only for the restricted language corresponding to normal logic programs
\cite{GRS91:WFS}.

The answer set semantics for extended logic
programs~\cite{GL90:ClassicalNegation} is defined via extensions which
are stable under a certain program transformation.  While this
semantics is a natural extension of stable
models~\cite{GL88:StableModels} and provides an elegant
model-theoretic semantics, there are several drawbacks which the
answer set semantics inherits from the stable models. In particular,
there is no efficient top-down proof procedure for the answer set
semantics, because the truth value of a literal $L$ may depend on the
truth value of a literal $L'$ which does {\em not} occur in the proof
tree below $L$~\footnote{See the extensive discussion
  in~\cite{AP96:WFSX} for details.}.  
The well-founded semantics~\cite{GRS91:WFS} is an
approximation of the stable model semantics, for which an efficient
top-down proof procedure exists. In~\cite{Prz90:ExtendedStable}, the
well-founded semantics is adapted to extended logic programs.
However, this semantics does not comply with the {\em coherence
principle}, which states that explicit negation implies implicit
negation.  In order to overcome this, \cite{PA92:WFSX,AP96:WFSX}
developed WFSX, a well-founded semantics for extended logic programs, 
which satisfies the coherence principle. It has several
desirable properties not enjoyed by the answer set semantics; in
particular, an efficient goal-oriented top-down proof procedure for
WFSX is presented in~\cite{ADP95:LPsystem}.  WFSX is well established
and e.g. widely available through Prolog implementations such as XSB
Prolog \cite{frssw97}.  

Our own work is complementary to these approaches, in that we fill a
gap by bringing argumentation and WFSX together in our definition of
$\na{u/a}$-justified arguments, which are equivalent to WFSX$_p$ 
\cite{Dam96:Thesis,AP96:WFSX,ADP95:LPsystem}, the
paraconsistent version of WFSX. Furthermore, the generality of our framework allows us to relate
existing argumentation semantics such as Dung's and Prakken and
Sartor's approach and thus provide a concise characterisation of all
the existing semantics mentioned above.

A number of authors \cite{KSE98:Argument,par96,sie97,PSJ98:AgentsArguing,stt01,tor02,Sch99:Argument,MA98:Argumentation}
work on argumentation for negotiating agents. Of these, the approaches
of \cite{stt01,tor02,Sch99:Argument} are based on logic programming.
The advantage of the logic programming approach for arguing agents 
is the availability of
goal-directed, top-down proof procedures. This is vital when
implementing systems which need to react in real-time and therefore
cannot afford to compute {\em all} justified arguments, as would be
required when a bottom-up argumentation semantics would be used.

In \cite{stt01,tor02}, abduction is used to define agent negotiation
focusing on the generation of negotiation dialogues using
abduction. This work is relevant in that it shows how to embed an
argumentation proof procedure into a dialogue protocol, which is
needed to apply proof procedures of argumentation semantics as defined
in this paper into agent communication languages such as
KQML~\cite{Fin94:KQML} or FIPA ACL~\cite{fipa}.

With a variety of argument-based approaches being pursued to define
negotiating agents, the problem of how these agents may inter-operate
arises. This paper could serve as a first step towards inter-operation
as existing approaches can be placed in our framework, thus making it
easier to compare them.

\section{Conclusion and Further Work}
\label{sec:conclusion}

We have identified various notions of attack for extended logic
programs. Based on these notions of attack, we defined notions of
acceptability and least fixpoint semantics. The contributions of this
paper are five-fold. 

\begin{itemize}
\item 
First, we defined a parameterised hierarchy of argumentation semantics
by establishing a lattice of justified arguments based on set
inclusion. We showed which argumentation semantics are equal, which
are subsets of one another and which are neither.

\item
Second, we examined some properties of the different
semantics, and gave a necessary and sufficient condition for a
semantics to satisfy the coherence principle~\cite{AP96:WFSX}, and a
sufficient criterion for a semantics to be consistent.

\item Third, we identified an argumentation semantics
  $J_{\na{u}/\na{a}}$ equal to the paraconsistent well-founded
  semantics for logic programs with explicit negation,
  WFSX$_p$~\cite{Dam96:Thesis,AP96:WFSX} and proved this equivalence.

\item
Forth, we established relationships between existing semantics, in
particular that $J_{Dung} \subsetneq J_{PS} \subsetneq
J_{\na{u}/\na{u}} = WFS \subsetneq J_{\na{u}/\na{a}} = WFSX_p$, where
$J_{Dung}$ and $J_{PS}$ are the least fixpoint argumentation semantics
of Dung~\cite{Dun93:ArgumentExplicit} and Prakken and
Sartor~\cite{PS97:Argument}, and $WFS$ is the well-founded semantics
without explicit negation~\cite{GRS91:WFS}.

\item Fifth, we have defined a
dialectical proof theory for argumentation.  For all notions of
justified arguments introduced, we prove that the proof theory is
sound and complete wrt. the corresponding fixpoint argumentation
semantics. 
\end{itemize}

It remains to be seen whether a variation in the notion of attack
yields interesting variations of alternative argumentation semantics
for extended logic programs such as preferred extensions or stable
extensions~\cite{Dun93:ArgumentExplicit}.  It is also an open question
how the hierarchy changes when priorities are added as defined
in~\cite{ant02,kak02,PS97:Argument,Bre96:DynamicPreferences,GSC98:Argument,%
Vre97:AbstractArgumentationSystems}.

\subsection*{Acknowledgement}
Thanks to Iara Carnevale de Almeida and Jos\'e J\'ulio Alferes for
fruitful discussions on credulous and sceptical argumentation
semantics for extended logic programming.\\
This work has been supported by EPSRC grant GRM88433.


\appendix

\section{Proofs of Theorems}
\label{appendix}

\setcounter{theorem}{2}

\begin{theorem}
  Let $x' \subseteq x$ and $y \subseteq y'$ be notions of attack, then
  $J_{x/y} \subseteq J_{x'/y'}$.
\end{theorem}

\begin{proof}
  We show by transfinite induction that 
  $J^\alpha_{x/y} \subseteq J^\alpha_{x'/y'}$, for all $\alpha$.

  \noindent {\it Base case}: $\alpha = 0$:
  Then $J_{x/y} = \emptyset = J_{x'/y'}$.

  \noindent {\it Successor ordinal}: $\alpha \leadsto \alpha + 1$:

  Let $A \in J^{\alpha+1}_{x/y}$, and $(B,A) \in x'$.
  Then also $(B,A) \in x$, and so there exists $C \in J^\alpha_{x/y}$
  such that $(C,B) \in y$, so also $(C,B) \in y'$.
  By induction hypothesis, $C \in J^\alpha_{x'/y'}$, 
  and so $A \in J^{\alpha+1}_{x'/y'}$.

  \noindent {\it Limit ordinal} $\lambda$:

  Assume $J_{x/y}^\alpha \subseteq J_{x'/y}^\alpha$ for all $\alpha < \lambda$.
  Then
  
  $J^\lambda_{x/y} = \bigcup_{\alpha < \lambda} J^\alpha_{x/y} \subseteq 
   \bigcup_{\alpha < \lambda} J^\alpha_{x'/y'} = J^\lambda_{x'/y'}$
\end{proof}

\setcounter{theorem}{3}
\begin{theorem}
  Let $x$ and and $y$ be notions of attack such that
  $x \supseteq \na{undercuts}$, and let
  $sy = y - \na{undercuts}^{-1}$.  
  Then $J_{x/y} = J_{x/sy}$.
\end{theorem}

\begin{proof} 
  By Theorem~\ref{thm:subset-just}, we have $J_{x/sy} \subseteq J_{x/y}$.
  We prove the inverse inclusion by showing that
  for all ordinals $\alpha$:
  $J_{x/y}^\alpha \subseteq J_{x/sy}^\alpha$,
  by transfinite induction on $\alpha$.\\[1ex]
  {\it Base case} $\alpha=0$:
  $J_{x/y} = \emptyset = J_{x/sy}$.\\[1ex]
  {\it Successor ordinal } $\alpha \leadsto \alpha+1$:
  Let $A \in J_{x/y}^{\alpha+1}$, and $(B,A) \in x$.
  By definition, there exists $C \in J_{x/y}^\alpha$ such that
  $(C,B) \in y$. By induction hypothesis, $C \in J_{x/sy}^\alpha$.

  If $B$ does not undercut $C$, then we are done. 
  If, however, $B$ undercuts $C$, then because 
  $C \in J_{x/sy}^\alpha$, and $\na{undercuts} \subseteq x$,
  there exists 
  $D \in J_{x/sy}^{\alpha_0} (\alpha_0 < \alpha)$ such that
  $(D,B) \in sy$. It follows that $A \in J_{x/sy}^{\alpha+1}$.\\[1ex]
  {\it Limit ordinal $\lambda$: }
  Assume $J_{x/y}^\alpha \subseteq J_{x/sy}^\alpha$ for all $\alpha < \lambda$.
  Then 
  $J_{x/y}^\lambda = \bigcup_{\alpha<\lambda}J_{x/y}^\alpha \subseteq
  \bigcup_{\alpha<\lambda}J_{x/sy}^\alpha = J_{x/sy}^\lambda$
\end{proof}

\setcounter{theorem}{5}
\begin{theorem}
  Let $x$ be a notion of attack such that 
  $x \supseteq$ \na{strongly attacks}. Then
  $J_{x/\na{u}} = J_{x/\na{d}} = J_{x/\na{a}}$.
\end{theorem}

\begin{proof}
  It is sufficient to show that $J_{x/\na{a}} \subseteq J_{x/\na{u}}$.
  Then by Theorem~\ref{thm:subset-just}, 
  $J_{x/\na{u}} \subseteq J_{x/\na{d}} \subseteq J_{x/\na{a}} = J_{x/\na{u}}$.\\[1ex]
%
%
  We prove by transfinite induction that for all ordinals $\alpha$:
  $J^\alpha_{x/\na{a}} \subseteq J^\alpha_{x/\na{u}}$.\\[2ex]
  {\it Base case: } $\alpha=0$\\[2ex]
  $J^\alpha_{x/\na{a}} = \emptyset =  J^\alpha_{x/\na{u}}$.\\[2ex]
  {\it Successor ordinal: } $\alpha \leadsto \alpha+1$\\[2ex]
  Let $A \in J_{x/\na{a}}^{\alpha+1}$, and $(B,A) \in x$.
  By definition, there exists $C \in J_{x/\na{a}}^\alpha$ such that
  $C$ undercuts or rebuts $B$. 
  By induction hypothesis, $C \in J_{x/\na{u}}^\alpha$.

  If $C$ undercuts $B$, then we are done. 
  If, however, $C$ does not undercut $B$, then $C$ rebuts $B$, 
  and so $B$ also rebuts $C$, i.e.\
  $B$ strongly attacks $C$.
  Because \na{strongly attacks} $\subseteq x$ and $C \in J_{x/\na{u}}^\alpha$,
  there exists 
  $D \in J_{x/\na{u}}^{\alpha_0} \subseteq J_{x/\na{u}}^\alpha$ ($\alpha_0 < \alpha$) 
  such that $D$ undercuts $B$. It follows that $A \in J_{x/\na{u}}^{\alpha+1}$.\\[2ex]
  {\it Limit ordinal $\lambda$: }\\[2ex]
  Assume $J_{x/\na{a}}^\alpha \subseteq J_{x/\na{u}}^\alpha$ for all $\alpha < \lambda$.
  Then 
  $J_{x/\na{a}}^\lambda = \bigcup_{\alpha<\lambda}J_{x/\na{a}}^\alpha \subseteq
  \bigcup_{\alpha<\lambda}J_{x/\na{u}}^\alpha = J_{x/\na{u}}^\lambda$.
\end{proof}

\setcounter{theorem}{6}
\begin{theorem}
  $J_{\na{sa}/\na{su}} = J_{\na{sa}/\na{sa}}$
\end{theorem}

\begin{proof}
  By Theorem~\ref{thm:subset-just}, 
  $J_{\na{sa}/\na{su}} \subseteq J_{\na{sa}/\na{sa}}$.

  We prove the inverse inclusion by showing that
  for all ordinals $\alpha$:
  $J_{\na{sa}/\na{sa}}^\alpha \subseteq J_{\na{sa}/\na{su}}^\alpha$,
  by transfinite induction on $\alpha$.\\[2ex]
  {\it Base case: } $n=0$\\[2ex]
  $J_{\na{sa}/\na{sa}}^0 = \emptyset = J_{\na{sa}/\na{su}}^0$\\[2ex]
  {\it Successor ordinal: } $\alpha \leadsto \alpha+1$\\[2ex]
  Let $A \in J_{\na{sa}/\na{sa}}^{\alpha+1}$, and $B$ strongly attacks $A$.
  By definition, there exists $C \in J_{\na{sa}/\na{sa}}^\alpha$ such that
  $C$ attacks $B$ and $B$ does not undercut $C$. 
  By induction hypothesis, $C \in J_{\na{sa}/\na{su}}^\alpha$.

  If $C$ undercuts $B$, then we are done. 
  If, however, $C$ rebuts $B$ and $C$ does not undercut $B$, 
  then $B$ also rebuts $C$, i.e.\
  $B$ strongly attacks $C$, and so because
  $C \in J_{\na{sa}/\na{su}}^\alpha$
  there exists 
  $D \in J_{\na{sa}/\na{su}}^{\alpha_0} \subseteq J_{\na{sa}/\na{su}}^\alpha$ ($\alpha_0 < \alpha$) such that
  $D$ strongly undercuts $B$. 
  It follows that $A \in J_{\na{sa}/\na{su}}^{\alpha+1}(\emptyset)$.\\[2ex]
  {\it Limit ordinal $\lambda$: }\\[2ex]
  Assume $J_{\na{sa}/\na{sa}}^\alpha \subseteq J_{\na{sa}/\na{su}}^\alpha$ for all $\alpha < \lambda$.
  Then 
  $J_{\na{sa}/\na{sa}}^\lambda = \bigcup_{\alpha<\lambda}J_{\na{sa}/\na{sa}}^\alpha \subseteq
  \bigcup_{\alpha<\lambda}J_{\na{sa}/\na{su}}^\alpha = J_{\na{sa}/\na{su}}^\lambda$.
\end{proof}

\setcounter{theorem}{7}
\begin{theorem}
  $J_{\na{su}/\na{a}} = J_{\na{su}/\na{d}}$
\end{theorem}

\begin{proof}
  By Theorem~\ref{thm:subset-just}, 
  $J_{\na{su}/\na{d}} \subseteq J_{\na{su}/\na{a}}$.\\[1ex]
%
%
  For the inverse inclusion, we show that
  for all ordinals $\alpha$:
  $J_{\na{su}/\na{a}}^\alpha \subseteq J_{\na{su}/\na{d}}^\alpha$,
  by transfinite induction on $\alpha$.\\[2ex]
  {\it Base case: } $\alpha=0$\\[2ex]
  $J_{\na{su}/\na{a}}^0 = \emptyset = J_{\na{su}/\na{d}}^0$\\[2ex]
  {\it Successor ordinal: } $\alpha \leadsto \alpha+1$\\[2ex]
  Let $A \in J_{\na{su}/\na{a}}^{\alpha+1}$, and $B$ strongly undercuts $A$.
  By definition, there exists $C \in J_{\na{su}/\na{a}}^\alpha$ such that
  $C$ undercuts or rebuts $B$. 
  By induction hypothesis, $C \in J_{\na{su}/\na{d}}^\alpha$.

  If $C$ undercuts $B$, or $B$ does not undercut $C$, then we are done. 

  Otherwise, $B$ strongly undercuts $C$, and so there exists
  $D \in J_{\na{su}/\na{d}}^{\alpha_0} \subseteq J_{\na{su}/\na{d}}^\alpha$ ($\alpha_0 < \alpha$) 
  such that $D$ defeats $B$.
  It follows that $A \in J_{\na{su}/\na{d}}^{\alpha+1}$.\\[2ex]
  {\it Limit ordinal $\lambda$: }\\[2ex]
  Assume $J_{\na{su}/\na{a}}^\alpha \subseteq J_{\na{su}/\na{d}}^\alpha$ for all $\alpha < \lambda$.
  Then 
  $$J_{\na{su}/\na{a}}^\lambda = \bigcup_{\alpha<\lambda}J_{\na{su}/\na{a}}^\alpha \subseteq
  \bigcup_{\alpha<\lambda}J_{\na{su}/\na{d}}^\alpha = J_{\na{su}/\na{d}}^\lambda$$
\end{proof}

\setcounter{theorem}{14}
\begin{lemma}
  Let $I$ be a two-valued interpretation.
  \begin{enumerate}
  \item 
    $L \in \Gamma(I)$ iff $\exists$ argument $A$ with conclusion $L$
    such that $\assm(A) \subseteq I$.
  \item 
    $L \in \Gamma_s(I)$ iff $\exists$ argument $A$ with conclusion $L$
    such that $\assm(A) \subseteq I$ and
    $\neg\conc(A) \cap I = \emptyset$.
  \item 
    $L \not\in \Gamma(I)$ iff $\forall$ arguments $A$ with conclusion $L$,
    $\assm(A) \cap I \not= \emptyset$.
  \item 
    $L \not\in \Gamma_s(I)$ iff $\forall$ arguments $A$ with 
    conclusion $L$, $\assm(A) \cap I \not= \emptyset$
    or $\neg\conc(A) \cap I \not= \emptyset$.
  \end{enumerate}
\end{lemma}
\begin{proof}\hspace{0em}
  \begin{enumerate}
  \item ``Only If''-direction:
    Induction on the length $n$ of the derivation of $L \in \Gamma(I)$.\\[0.5ex]
    \textit{Base case}: $n = 1$:\\
    Then there exists a rule $L \gets not~L_1,\ldots,not~L_n$ in $P$ s.t.\
    $L_1,\ldots,L_n \not\in I$,
    and $[L \gets not~L_1,\ldots,not~L_n]$ is an argument for $L$ 
    whose assumptions are contained in $I$.\\[0.5ex]
    \textit{Induction step}: $n \leadsto n+1$:\\
    Let $L \in \Gamma^{n+1}(I)$. Then there exists a rule 
    $r = L \gets L_1,\ldots,L_n,not~L'_1,\ldots,L'_m$ in $P$ s.t.\
    $L_i \in \Gamma^n(I)$, and $L'_i \not\in I$.
    By induction hypothesis, there exists arguments $A_1,\ldots,A_n$ for
    $L_1,\ldots,L_n$ with $\assm(A_i) \subseteq I$.
    Then $A = [r] \cdot A_1 \cdots A_n$ is an argument for $L$
    such that $\assm(A) \subseteq I$.\\[1ex]
    ``If'' direction: Induction on the length of the argument.\\[0.5ex]
    \textit{Base case}: $n = 1$:\\
    Then $A = [L \gets not~L_1,\ldots,not~L_n]$, and 
    $L_1,\ldots,L_n \not\in I$. Then $L \gets \in \frac{P}{I}$,
    and $L \in \Gamma^1(I)$.\\[0.5ex]
    \textit{Induction step}: $n \leadsto n+1$:\\
    Let $A = [L \gets L_1,\ldots,L_n,not~L'_1,\ldots,not~L'_m;r_2,\ldots,r_n]$
    be an argument s.t.\ $\assm(A) \subseteq I$.
    $A$ contains subarguments $A_1,\ldots,A_n$ for $L_1,\ldots,L_n$,
    with $\assm(A_i) \subseteq I$. 
    Because $L'_1,\ldots,L'_m \not\in I$, then
    $L \gets L_1,\ldots,L_n \in \frac{P}{I}$.
    By induction hypothesis, $L_i \in \Gamma(I)$.
    so also $L \in \Gamma(I)$.
  \item ``Only If''-direction:
    Induction on the length $n$ of the derivation of $L \in \Gamma_s(I)$.\\[0.5ex]
    \textit{Base case}: $n = 1$:\\
    Then there exists a rule $L \gets not~L_1,\ldots,not~L_n$ in $P$ s.t.\
    $\neg L,L_1,\ldots,L_n \not\in I$,
    and $[L \gets not~L_1,\ldots,not~L_n]$ is an argument for $L$ 
    whose assumptions are contained in $I$, and $\neg L \not\in I$.\\[0.5ex]
    \textit{Induction step}: $n \leadsto n+1$:\\
    Let $L \in \Gamma^{n+1}(I)$. Then there exists a rule 
    $r = L \gets L_1,\ldots,L_n,not~L'_1,\ldots,L'm$ in $P$ s.t.\
    $L_i \in \Gamma^n(I)$, $L'_i \not\in I$, and $\neg L \not\in I$.
    By induction hypothesis, there exists arguments $A_1,\ldots,A_n$ for
    $L_1,\ldots,L_n$ with $\assm(A_i) \subseteq I$ and 
    $\neg \conc(A_i) \cap I = \emptyset$.
    Then $A = [r] \cdot A_1 \cdots A_n$ is an argument for $L$
    such that $\assm(A) \subseteq I$, and $\neg \conc(A) \cap I = \emptyset$.\\[1ex]
    ``If'' direction: Induction on the length of the argument.\\[0.5ex]
    \textit{Base case}: $n = 1$:\\
    Then $A = [L \gets not~L_1,\ldots,not~L_n]$, and 
    $\neg L,L_1,\ldots,L_n \not\in I$. Then $L \gets \in \frac{P_s}{I}$,
    and $L \in \Gamma^1(I)$.\\[0.5ex]
    \textit{Induction step}: $n \leadsto n+1$:\\
    Let $A = [L \gets L_1,\ldots,L_n,not~L'_1,\ldots,not~L'_m;r_2,\ldots,r_n]$
    be an argument s.t.\ $\assm(A) \subseteq I$, and 
    $\neg \conc(A) \cap I = \emptyset$.
    $A$ contains subarguments $A_1,\ldots,A_n$ for $L_1,\ldots,L_n$,
    with $\assm(A_i) \subseteq I$, and $\neg \conc(A_i) \cap I = \emptyset$. 
    Because $L'_1,\ldots,L'_m \not\in I$, and $\neg L \not\in I$, then
    $L \gets L_1,\ldots,L_n \in \frac{P}{I}$.
    By induction hypothesis, $L_i \in \Gamma(I)$,
    so also $L \in \Gamma(I)$.
  \item and 4. follow immediately from \ref{lemma:model-arg:u}.\ and
    \ref{lemma:model-arg:a}.\ because $I$ is two-valued.
  \end{enumerate}
\end{proof}

\setcounter{theorem}{18}
\begin{theorem}
  An argument $A$ is $x/y$-justified iff there exists a $x/y$-dialogue
  tree with $A$ as its root, and won by the proponent.
\end{theorem}

\begin{proof}
  ``If''-direction.
  We show by transfinite induction: If $A \in J^{\alpha}_{x/y}$, then 
  there exists a winning $x/y$-dialogue tree of height $\leq \alpha$ 
  for $A$.\\[1ex]
  {\em Base case $\alpha=0$:}\\
  Then there exists no argument $B$ such that $(B,A) \in x$,
  and so $A$ is a winning $x/y$-dialogue tree for $A$ of height $0$.\\[0.5ex]
  {\em Successor ordinal $\alpha+1$:}\\
  If $A \in J_{x/y}^{\alpha+1}$, then for any $B_i$ such that $(B_i,A) \in x$
  there exists a $C_i \in J_{x/y}^{\alpha}$ such that $(C_i,B_i) \in y$.
  By induction hypothesis, there exist winning $x/y$-dialogue trees
  for the $C_i$. Furthermore, if any of the $C_i$ contains a move 
  $m = (P, A)$, then it also contains a winning subtree for $A$ rooted
  at $m$ and we are done. Otherwise, we have a winning tree rooted at $A$,
  with children $B_i$, whose children are the winning trees for $C_i$.\\[0.5ex]
  {\em Limit ordinal $\lambda$:}\\
  If $A \in J_{x/y}^{\lambda}$, then there exists an $\alpha < \lambda$
  such that $A \in J_{x/y}^{\alpha}$; by induction hypothesis,
  there exists a winning $x/y$-dialogue tree of height $\alpha$ for $A$.\\[1ex]
  ``Only-if''-direction.
  We prove by transfinite induction: 
  If there exists a winning tree of height $\alpha$ for $A$,
  then $A \in J_{x/y}^{\alpha}$.
  
  Note that by definition, the height of a dialogue tree is
  either $0$ or a successor ordinal $\alpha+1$. So we prove the base
  case $0$, and for the induction step, we assume that the induction
  hypothesis holds for all $\beta < \alpha+1$.\\[1ex]
  {\em Base case $\alpha=0$:}\\
  Then there are no arguments $B$ such that $(B,A) \in x$, and so
  $A \in J_{x/y}^0$.\\[0.5ex]
  {\em Successor ordinal $\alpha+1$:}\\
  Let $T$ be a tree with root $A$, whose children are $B_i$,
  and the children of $B_i$ are winning trees rooted at $C_i$.
  By induction hypothesis, $C_i \in J_{x/y}^{\alpha}$.
  Because the $B_i$ are all those arguments such that $(B_i,A) \in x$,
  then $A$ is defended against each $B_i$ by $C_i$, and so
  $A \in J_{x/y}^{\alpha+1}$.
\end{proof}

\end{document}